\documentclass[journal=jpcafh, manuscript=article, layout=twocolumn]{achemso}

\usepackage{amsmath, amssymb}%
\usepackage{subcaption}%
\usepackage{multirow}
\usepackage[version=4]{mhchem}%

\usepackage{xr-hyper}%
\usepackage{xr}%

\makeatletter
\newcommand*{\addFileDependency}[1]{
  \typeout{(#1)}
  \@addtofilelist{#1}
  \IfFileExists{#1}{}{\typeout{No file #1.}}
}
\makeatother

\usepackage{color}

\title{
Limitations of atomistic modeling to reveal ejection of proteins
from charged nanodroplets
}

\author{Victor Kwan}
\affiliation{
Department of Chemistry, The University of Western Ontario, London, Ontario, Canada N6A 5B7
}

\author{Pranav Ballaney}
\affiliation{
Department of Chemistry, The University of Western Ontario, London, Ontario, Canada N6A 5B7
}

\author{Titiksha}
\affiliation{   
Department of Chemistry, The University of Western Ontario, London, Ontario, Canada N6A 5B7
}

\author{Styliani Consta}
\email{sconstas@uwo.ca}
\affiliation{
Department of Chemistry, The University of Western Ontario, London, Ontario, Canada N6A 5B7
}

\begin{document}

\begin{abstract}
Molecular dynamics using atomistic
modeling is frequently used
to unravel the mechanisms of macroion release from electrosprayed droplets.
However, atomistic modeling is currently feasible for only the smallest
window of droplet sizes appearing at the end of a disintegrating droplet's lifetime.
The relevance of the observations made to the actual
droplet evolution, which is much longer that the simulated
sizes, has not been addressed.
Here, we perform a systematic study of desolvation mechanisms of 
poly(ethylene glycol) (PEG), protonated peptides of
different compositions and proteins in 
order to (a) examine whether atomistic modeling can 
establish the extrusion mechanism of proteins from droplets and
(b) obtain insight for the charging mechanism in larger droplets than
those simulated.
Atomistic modeling of PEG charging shows that above a critical droplet
size charging occurs transiently by transfer of ions from the solvent to
the macroion, while below the critical size, the capture of the ion
from PEG has a lifetime sufficiently long for extrusion of the charged PEG from
an aqueous droplet. This is the first report of the role of  droplet curvature in the
conformations and charging of macroions. 
Simulations with highly hydrophobic peptides show that
partial extrusion of a peptide from the droplet surface is
rare relative to desolvation by drying-out.
Differently from what has been presented in the
literature we argue that atomistic simulations have not sufficiently established extrusion
mechanism of proteins from droplets and their charging
mechanism.
Moreover, we argue that release of highly charged proteins can occur at an
earlier stage of a droplet's lifetime than predicted by atomistic modeling.
In this earlier stage, we emphasize the key role of jets emanating from a droplet at the point
of charge-induced instability in the release of proteins. 

\end{abstract}

\section{Introduction}
\label{sec:Intro}

\begin{figure*}
        \centering
        \includegraphics[width=\textwidth]{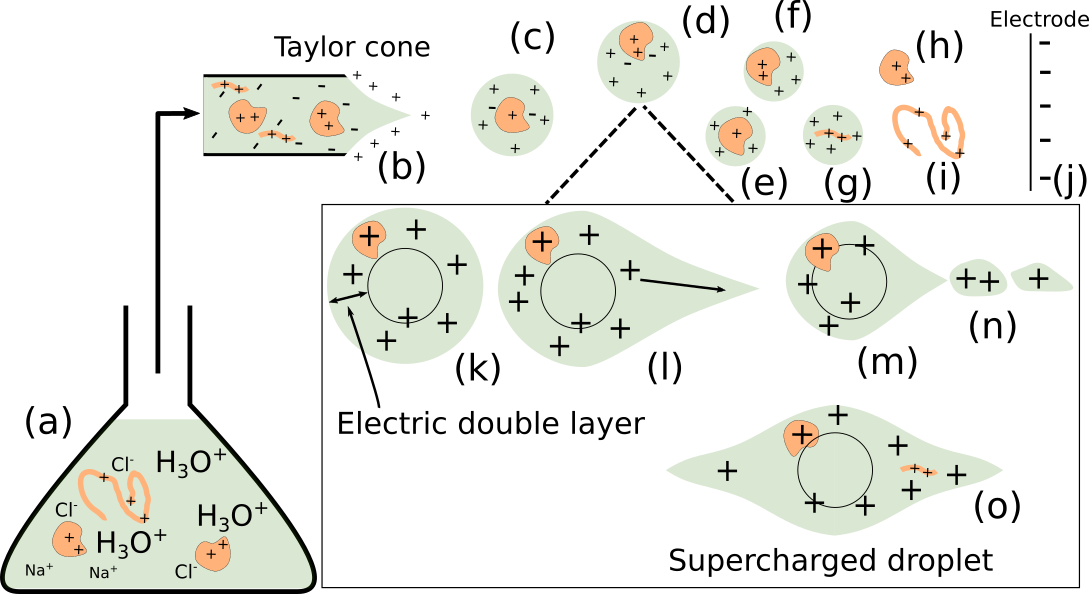}
        \caption{Schematic representation of transfer of macroions 
	  from the bulk solution to droplets to the gaseous phase in an electrospray process.
	  Macroions are colored orange
	  and can be compact or extended\cite{constacpl2016}.
	  The ``+'' and ``-'' signs indicate charge carriers 
	  such as \ce{Na+}, \ce{NH4+}, \ce{H3O+} for ``+'', and \ce{CH3COO-}, \ce{Cl-} for ``-''.
	(a) Macroions in bulk solution that is to be sprayed. 
	The ion distribution around them is described by
      Manning-Oosawa model\cite{kirkwood1952forces, oosawa1971}. 
      The solution is transferred to a capillary. (b) A Taylor cone\cite{taylor1964, fernandezdelamora2007} 
      at the
      tip of the capillary formed due to an applied external 
      electric field. The polarity of the electric field can be positive or
      negative. In the schematic the electrospray is in the positive ion mode 
      as indicated by the negative electrode in (j). 
      (c)-(d) Nascent droplet aerosol originating from the
      Taylor cone. The droplets contain macroions, excess co-ions and
      counterions.  The macroions can be
    found in different locations in the droplets. 
      (e)-(g) Droplets travel along the electric field and shrink by undergoing
	a repeating cycle of solvent evaporation and Coulomb explosion events.
      (h)-(i) In the last stage desolvated macroions emerge.  (j) Negative electrode that sets 
      the polarity of the applied external electric field.
    (k)-(o) Details of a droplet's division mechanism 
    via jet formation \cite{consta2022atomistic}. The circle within (k)-(o) 
    indicates the inner boundary of the electric double 
    layer\cite{constantopoulos1999effects, enke1997predictive, 
    kwan2020bridging, kwan2020molecular, kwan2021relation}. In 
    previous research\cite{kwan2020bridging, kwan2020molecular, kwan2021relation} 
    we estimated its thickness to be 1.5~nm-2.0~nm regardless of droplet size.
    Variations of external pressure along a droplet's journey are not
    shown. Details are discussed in the text.}
\label{fig:esi-schematic}
\end{figure*}

The chemical processes within ionization techniques, employed to
transfer analyte species
from the bulk solution into the
gaseous phase for mass spectrometry (MS) analysis, have always been challenging to 
investigate. 
The question of ``what is the mechanism by which a macroion obtains its charge?'' 
appears with the first successes of the methods\cite{dole1968molecular, mack1970molecular, 
fenn1993ion, delamora2000, loscertales1995experiments, 
karas1988laser, trimpin2017spontaneous, trimpin2018vacuum, 
apsokardu2021ion, pervukhin2020aerodynamic, martini2021splashing, finlaysonpitts2022experiments,
grimm2009evaporation, Smith2002, upton2017easily, jarrold2021applications}.

In spray-based ionization
techniques that include sonic\cite{hirabayashi1994sonic}, thermal\cite{blakley1983thermospray}, and 
electrospray\cite{dole1968molecular, fenn_88}, droplets are the vehicles
that transfer analytes from the bulk solution into the gaseous phase.
Regardless of the spraying method, the droplets are charged
and they are composed of solvent, which is often water,
simple charge carriers such as \ce{H3O+}, \ce{Na+} ions and the charged macroion.
During their lifetime, they disintegrate via solvent evaporation and Coulomb fission
events. 

The droplet environment plays a decisive role in a macroion's charge state.
It is expected that competition between the dynamics of 
various processes such as solvent evaporation, proton transfer reactions\cite{zare2016, wei2020accelerated}, 
Coulomb fission, 
possible conformational changes of
macroions 
will determine a macroion's charge state. 
Because of the multitude of processes that take place in a droplet's lifetime,
the precise mechanism via which macroions obtain their charge has not
been settled yet. However, fundamental mechanisms have been 
proposed\cite{delamora2000, enke1997predictive, fenn1993ion, fenn1997electrospray, constapeg1,
consta2012}, which 
are analyzed in the next section.

Molecular dynamics (MD) using atomistic modeling is broadly used to reveal
the macorion release mechanisms. However, atomistic modeling is limited to monitor
disintegration of droplets comprised at most a couple of thousands of \ce{H2O} 
molecules, a protein and ions. This system size corresponds to an equimolar radius of 
several nanometers.
In atomistic modeling it has been considered that macroion release occurs in this narrow window in
a droplet's lifetime without considering its long history. The problem is more severe
at elevated temperature where non-equilibrium conditions are favored and thus, the droplet's
history plays a decisive role. Here, we addressed the feasibility of ejection mechanisms
of macroions from charged aqueous nanodroplets using atomistic modeling. 
Poly(ethylene glycol) (PEG), protonated peptides of varying
degree of hydrophobicity and myoglobin are selected as typical examples to
demonstrate the likelihood of ejection.
We also examine, how the insight obtained from the atomistic modeling of the
smallest nanodroplets is
transferable to droplets of larger size.

In order to provide the broader context of the problem and the limitations
of the atomistic modeling in Fig.~\ref{fig:esi-schematic} we present a schematic of the 
entire aerolization process from the bulk solution to the
release of macroions. 
In the bulk solution polyelectrolytes (macroions) (Fig.~\ref{fig:esi-schematic}~(a))
are surrounded by a cloud of counterions, 
which affects their radius of gyration and 
conformation\cite{chremos2016counter, dobrynin2005theory, kirkwood1952forces, oosawa1971}. 
The solution is injected to a capillary (Fig.~\ref{fig:esi-schematic}~(b))
at the tip of which there is an external electric field.
The application of the electric field induces  
a conical protrusion in the liquid surface that is called 
the Taylor cone\cite{taylor1964, fernandezdelamora2007} 
shown in Fig.~\ref{fig:esi-schematic}~(b).
This is a critical step in the ionization process, where the liquid-vapor interface of the cone 
accumulates ions of the same sign depending on the polarity of the 
applied electric field.
The cone emits charged droplets
as shown in Fig.~\ref{fig:esi-schematic}~(c)-(d), where the charge carriers
are the macroions and simple ions such as \ce{H3O+}, \ce{Na+}, \ce{NH4+} ions. Counterions
such as \ce{Cl-}, \ce{CH3OO-} are also present in the nascent aerosol. 
The dimensions of the initially produced
droplets depend on
the specifics of the instrument and usually vary between a few hudrends
of nanometers to micrometers.
It is noted that in a huge
ensemble of sprayed droplets a macroion
can be statistically found anywhere in a droplet's volume as shown in 
Fig.~\ref{fig:esi-schematic}~(c)-(d).
However, because of the prevalence of vapor-liquid interface, the likelihood for a macroion
to be found on the surface is considerably higher than in the interior. 
It is believed
that the macroions maintain the surrounding cloud of counterions as in the
bulk solution\cite{wilm2011principles}. 
Even though this assumption is reasonable, we think that
it may break down because of the preference of macroions
to be located
in the outer layers of droplets where they may be
affected by the electric double layer
\cite{kwan2020bridging, kwan2020molecular, kwan2021relation}. 
This question is investigated in future work.

The sprayed droplets undergo a cycle of solvent evaporation and
ion ejection events (Fig.~\ref{fig:esi-schematic}~(c)-(g)), until the macroions
(linear or compact) emerge (Fig.~\ref{fig:esi-schematic}~(h)-(i)).
For example, the droplets shown in Fig.~\ref{fig:esi-schematic}~(c)-(d) may decrease 
in size by solvent evaporation.
Once a droplet's size reaches a critical point at which the electrostatic repulsive 
forces among excess ions of the same sign
overcome the surface tension forces, the droplet divides.
The point where these two forces are equal is called the 
Rayleigh limit\cite{rayleigh1882, hendricks1963, peters1980rayleigh, 
consta2015disintegration, oh2017droplets} (RL).

The RL is defined via the Rayleigh fissility parameter ($X$) given by 
\begin{equation}
  \label{eq:fissility}
  X = \frac{ Q^2} {64 \pi ^2 \gamma \varepsilon _0 R^3}
\end{equation}
where $Q$ is the droplet charge, $\gamma$ the surface tension, 
$\varepsilon_0$ and $R$ are the permittivity of vacuum and the radius of the
droplet, respectively. 
When $X = 1$, the charged droplet is at the RL.
At $X<1$ (``below'' the RL) the droplet is stable wrt to small perturbations of its
spherical shape.
At $X>1$ (``above'' the RL) the droplet is unstable.

The mechanism by which a droplet emits ions is shown in 
Fig.~\ref{fig:esi-schematic}~(k)-(o).
The circle in Fig.~\ref{fig:esi-schematic}~(k) indicates the presence of an electric
double layer\cite{constantopoulos1999effects, enke1997predictive, 
    kwan2020bridging, kwan2020molecular, kwan2021relation}.

At the RL a single jet is formed similar to a Taylor cone (Fig.~\ref{fig:esi-schematic}~(l)),
which may provide the path for single ions and macroions to be emitted\cite{consta2022atomistic} (Fig.~\ref{fig:esi-schematic}~(n)). 
An important result
of our previous research is that a pronounced conical fluctuation precedes the  
transfer of the ions within the conical region\cite{consta2022atomistic} (Fig.~\ref{fig:esi-schematic}~(l)).
Experiments have often observed two cones instead of one\cite{gomez1994charge, duft2003}.
We have found that rapid solvent evaporation
may lead to supercharged droplets (above the RL) 
that may form two jets in diametrically opposite sites\cite{kwan2021relation} as depicted
in Fig.~\ref{fig:esi-schematic}~(o). The multiple cones may enhance the probability of 
macroion ejection in these larger droplets.

In the nascent sprayed ensemble of droplets, the majority
of the macroions will be found on the droplet surface, thus, they will be
more susceptible to ejection either via jets or other droplet 
fluctuations appearing under highly non-equilibrium conditions that
the droplets may be found.
Thus, atomistic modeling of the minute nanoscopic droplets does not account for 
a number of possible macroion charging mechanisms as it is discussed in 
detail in the next section.

Here, we perform a systematic study to examine whether ejection of an
unstructured linear macroion such as PEG and proteins is a dominant
mechanism of release in droplets composed of a few thousands \ce{H2O}
molecules and the possible charging mechanisms. 
We consistently find that ejection of a protonated peptide and protein is of low likelihood 
in droplets composed up to a $2.3 \times 10^4$
\ce{H2O} molecules (equimolar radius of 5.6~nm). By elimination, 
we suggest, that any possible ejection
observed in experiments may come from capturing of a macroion into
Rayleigh jets.
We also identify three new factors that enter the charging of linear
macroions that have not been recognized in earlier studies: (a) 
Effect of droplet curvature in the conformation of a linear 
macroion; (b) Role of conical shape
fluctuations in low probability events
of macroion ejection; and (c) Partial extrusion of a macroion near the RL may be
viewed as a Taylor cone.

\section{A critical view of macroion release from droplets}

\begin{figure*}
        \centering
        \includegraphics[width=\textwidth]{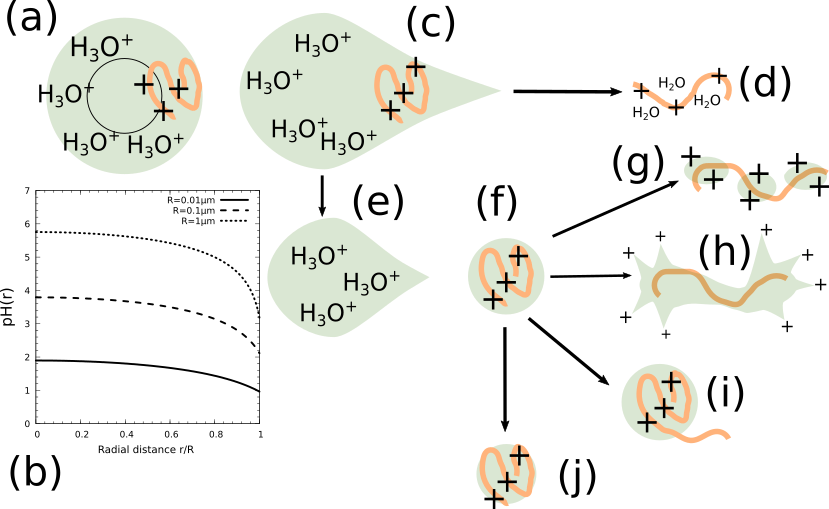}
        \caption{Schematic representation of the process of transfer of linear proteins
	  from a charged droplet near the Rayleigh limit
	  to the gaseous phase in an electrospray process.
	  The color coding and the meaning of charge signs
	  are the same as in Fig.~\ref{fig:esi-schematic}.
	  (a) Charged droplet near the Rayleigh limit. 
	  (b) pH profile as a function of the distance from a droplet's center
	  of mass (COM) for droplets of various sizes from the solution of the
	  Poisson-Boltzmann equation for a spherical geometry\cite{consta2013acid, chamberlayne2020simple}.
	  In the x-axis $r$ denotes the distance from the droplet's COM and
	  $R$ the droplet's radius.
	  (c) Rayleigh jet formation
	  on the droplet surface that may capture an entire chain or a segment of it.
	  (d) A solvated chain may be entirely ejected from the jet or a segment may be ejected. 
	  (e) The jet may
	  release a droplet containing the macroion. It is very likely that additional 
	  simple ions will not be released in the droplet containing
	  the macroion. (f) The droplet may take ``pearl-necklace''
	  conformations\cite{consta2010manifestation}. (g) 
	  The droplet may become ``star-shaped\cite{consta2010manifestation}''.
	  (h) The droplet may gradually release a linear macroion\cite{consta2010manifestation, 
	  constapeg1, constapeg2, constapeg2015}.
	  (i) The droplet may shrink and a protein (compact or linear) 
	  may transfer protons to the surrounding
	  \ce{H2O} molecules\cite{consta2013acid}.
Details of all the steps are discussed in the text.}
\label{fig:senarios}
\end{figure*}

An early review by Kebarle\cite{kebarle1993ions} describes
the initial efforts in the literature toward understanding the mechanisms of macroion charging and release
from droplets.
In Ref.~\cite{kebarle1993ions} two mechanisms of linear 
protein release are described.
One of them was proposed by J. Fenn, who suggests that proteins or other analytes desorb from droplets by capturing 
an amount of charge that is covered by their area, which is in contact with
the droplet surface\cite{fenn1993ion}. According to this
conjecture, a protein exposes a charged portion on the
surface, this part is extruded from the surface similar to the manner that a single ion
escapes in the ion evaporation mechanism\cite{iribarne1976, iribarne1979, kwan2022conical} (IEM), and finally the extruded part pulls out the rest of the chain
until the chain is desorbed\cite{kebarle2000brief}. Fenn was using the term ``ion desorption
mechanism'' (IDM) for the ejection of macroions from droplets in analogy
to IEM \cite{nguyen2007, fenn1993ion, fenn1997electrospray}. 
Another mechanism was proposed by P. Kebarle, where a protein found on a droplet's surface leaves
via a significant oscillation of the droplet shape that expels a progeny droplet containing the protein\cite{kebarle1993ions}.
In both propositions, it is expected that the protein is found near the vapor-droplet
interface. Considering that the ensemble of droplets has a high proportion of interface,
the likelihood of a protein to be near the interface is substantial. 
The conducting nature of the droplet is another important force that 
drives a protein near the surface. 
It is noted that the Fenn and Kebarle propositions do not clearly define the droplet size from where 
the proteins are released. Nevertheless, Kebarle's mechanism implies that the 
droplet has to be substantially larger than the protein in order to undergo 
such a large deformation. Fenn's mechanism does not consider the role of 
the droplet's shape fluctuations, such as the conical fluctuations we have identified and the 
the Rayleigh jets\cite{consta2022atomistic} that they may play a role in a macroion's charging
and release.
Both mechanisms are captured in continuum theory we have developed\cite{consta2012, constacpl2016}, since 
atomistic modeling is limited in minute nanodroplets.
The charging of PEG is an excellent example that can be used to test Fenn's proposition
in minute nanodroplets. A systematic change of the macroion and droplet size may allow us to
extrapolate the results in larger droplets.

In 2011, for the first time we presented direct 
evidence of the ejection\cite{constapeg1} of linear macroions from
droplets by using atomistic modeling of the sodiation (or lithiation) of 
poly(ethylene glycol) (PEG). 
We studied the sodiation of PEG within different solvents 
(\ce{H2O}, \ce{CH3CN}, \ce{CH3OH}) and in water with different 
ions (\ce{Na+}, \ce{Li+}, \ce{Ca^2+}).
The PEG charging and extrusion mechanism have been tested with different force fields
and simulation
codes and the results show remarkable 
consistency\cite{constapeg1, constapeg2, constapeg2015, oh2017charging, InOhPHDthesis}.

The similarities between PEG and proteins have been addressed by de la Mora who notes\cite{delamora2000}
``Proteins have been by far the species most studied
by ESMS. However, most related data have involved
denaturing conditions, with geometrical ambiguities
similar to those with PEGs, both for the ion and
for the drops producing them.''
For proteins the direct
protonation cannot be observed yet in atomistic modeling because of the complexity of the interactions
that require quantum chemistry modeling,
long time
scale for conformational changes that may lead to a possible extrusion and strong dependence of the 
phenomena on force field parameters\cite{zapletal2020choice}.

Konermann et al. using atomistic modeling reported 
that myoglobin in charge states $+17$ and $+23$ is fully ejected from droplets composed of
$2.0 \times 10^4-2.3 \times 10^4$~\ce{H2O} molecules and they named it the 
chain-ejection mechanism\cite{konermann2013, konermann2012} (CEM). 
CEM hypothesizes the 
charging of the macroion in a manner
identical to that of the PEG sodiation (or lithiation), but with an imaginary proton in the position
of \ce{Na+}. The proton does not have any
atomistic presence in the modeling and does not transfer on the chain as \ce{Na+} or \ce{Li+}
do in PEG.
CEM does not account for the fact that there 
are different chain extrusion mechanisms depending on the solvation energy
of the chain\cite{constacpl2016}.

Results from our previous research 
lead us to synthesize the picture shown in
Fig.~\ref{fig:senarios} about the charging and release of macroions from 
droplets larger than those that
can be modeled atomistically.  
Figure~\ref{fig:senarios} presents in more detail mechanisms of protein charging and release
in the region shown in Fig.~\ref{fig:esi-schematic} (c)-(g).

There are two possible charging and release mechanisms of proteins 
(linear or compact) that cannot be captured by the atomistic modeling of the minute
nanodroplets but have been inferred from continuum modeling\cite{consta2012} and by the atomistic
modeling of the smallest jet\cite{consta2022atomistic}. 

\paragraph{Charging of a linear protein with the assistance of the
jet} Differently from the conducting body of a charged droplet,
    a conical jet has electric field that attracts ions. It is likely that
    a segment of a linear protein near the surface is caught in
    the cone and the transfer of the \ce{H3O+} ions in this region
    facilitate the charging of the protein. The cone may lead to
    a partial or complete extrusion of a linear protein.

\paragraph{Protonation of a protein in large droplets and emission from a jet}
Charging of a protein (linear or compact) can occur in the outer low pH 
    layer of droplets with radius of at least a few tens of nanometers.
    We have shown (Fig.~\ref{fig:senarios}~(b)) that depending of the droplet radius,
    an outer layer of considerable thickness has significantly lower
    pH than the bulk of the droplet. The likelihood for a protein
    to be in this layer is high because of co-operating energetic and 
    entropic factors. Once the protein is charged
    it is likely to be captured in a Rayleigh jet (Fig.~\ref{fig:senarios}~(c)-(d)). 
   In previous research we have estimated a jet angle of $20^\circ$
    in an aqueous droplet. If the base of the jet is approximately the size of the
    protein, then we can estimate the radius $R_{emitted}$ of the emitted droplet  
    containing the protein, which will be $R_{emitted} = R_P + 0.33 R_P$,
    where $R_P$ is the radius of the protein. Thus, a compact protein will be
    released surrounded by one or two layers of \ce{H2O}, while a linear protein
    will be solvated with the corresponding amount of \ce{H2O} along
    its backbone. In larger droplets the base of the cone can be larger than
    the protein size, thus more \ce{H2O} layers can surround the protein.
    Once the droplet is emitted, if the protein unfolds then it may form pearl-necklace 
    conformations (Fig.~\ref{fig:senarios}~(g)),
    star-shapes (Fig.~\ref{fig:senarios}~(h)) or it may show a gradual 
    extrusion (Fig.~\ref{fig:senarios}~(i)) as we have discussed in 
    previous work\cite{constapeg2015}.
    A mechanism where a protein in a small droplet can release charge has been
    discussed in previous work\cite{consta2013acid}.

Regarding, the capture of the protein in a jet, one may argue that since there are at most two proteins (or protein complexes) in a 
  droplet, the likelihood
to be caught in a jet is low because (i) there are many more simple ions than a protein, and (ii) 
in general lack of correlation between the ion location and jet formation\cite{consta2022atomistic,
kwan2021relation, kwan2022conical}. However, the 
likelihood may increase because in slightly supercharged droplets multi-jets may appear  
\cite{consta2022atomistic,
kwan2021relation, kwan2022conical} and also, a loosely structured protein may occupy
a significant area on a droplet's surface that may increase the likelihood for a segment of it to be
caught in a jet.

Temperature will play an important role in the charge state and release of proteins. 
The temperature of droplets has not been established 
yet\cite{soleilhacantoine2015, gibsoncook2013, Smith2002, grimm2009evaporation} 
partly because it will depend on the details of the instrument and
the specific experiment. Factors that affect the temperature of
microdroplets, whose temperature has been mostly studied\cite{soleilhacantoine2015, gibsoncook2013}, are
evaporative cooling, conductive thermal transfer with the sheath gas, friction with the background gas.
Temperature measurements that have been reported for microdroplets are in the 
range of 270~K to 307~K.
For speeding up solvent evaporation atomistic modelling has been performed in very
elevated temperature of 370~K or higher that may not be representative
of experimental conditions. Methods of effective treatment of
solvent evaporation at room temperature have been reported\cite{consta2022atomistic}.

The various scenarios of protein charging and release
can lead to the charge state of proteins because in all the scenarios the proteins are found 
in a highly acidic environment. 

\section{Systems and Simulation Methods}

Molecular dynamics (MD) simulations of aqueous charged droplets which also include
poly(ethylene glycol) (PEG),
protonated peptides, protonated myoglobin and \ce{Na+} ions 
were performed by using the software NAMD version 2.12\cite{Phillips2005}. For comparison with
other works in the literature,
simulations of myoglobin in certain charge states were also performed with GROMACS 2018.3 
\cite{hess2008, berendsen1995gromacs, abraham2015gromacs}. 
Details of the set-up
are discussed in the next paragraphs for every macroion.
Visualization of all the trajectories was performed by using VMD 1.9.2\cite{Humphrey1996}.

\subsection{PEG and protonated peptides}
In order to examine the likelihood of ejection of unstructured
linear macroions, we selected poly(ethylene glycol) (PEG), and protonated peptides with
a variable degree of hydrophobicity as representative examples.
The systems and conditions are described in Table~\ref{table:peg-pept}.

PEG was composed of 54 monomers (PEG54) and was embedded in droplets
composed of $5\times 10^3$ and $10\times 10^3$ \ce{H2O} molecules. 
A number of \ce{Na+} ions were added so as the systems are near the RL.
PEG54 is long enough to capture
up to a maximum of  4\ce{Na+} ions\cite{constapeg1, constapeg2}. We selected a relatively short PEG for which is
feasible to directly simulate its charging in aqueous droplets considerably larger than its size.
This realistic system will allow us to infer charging mechanisms in droplets much larger
than those that can be atomistically modeled so far.
The structure of PEG54 and force field parameters were generated using 
CHARMM-GUI\cite{Choi2021CHARMM,Jo2008CHARMM,Lee2015CHARMM}. 
The water molecules were modeled with the modified-TIP3P (mTIP3P) model. 

We also prepared protonated peptides with hydrophobic segments composed of valine (Val) residues and 
a few single charged lysine (\ce{Lys^+}) residues  in between these segments.
Different arrangements of valine segments and \ce{Lys^+} were tested.
We performed equilibrium and non-equilibrium MD simulations on 
three model protonated peptide sequences
\ce{(Val_{6}-Lys^+)_3-Val_{6}}, \ce{(Val_{10}-Lys^+)_3} and \ce{Val_{10}-Lys^+-Val_{10}} embedded
in sodiated aqueous droplets. 
The prevalence of hydrophobic segments is intended to 
increase the peptide's propensity for the droplet's surface and
thus, make it more prone to ejection.
The protonated peptide was solvated in $\approx 2100$ \ce{H_2O} molecules, and 11 \ce{Na+} ions 
were added to bring the total charge of the system to 14 $e$ for \ce{(Val_{6}-Lys^+)_3-Val_{6}} 
and \ce{(Val_{10}-Lys^+)_3-Val_{10}} and 12 $e$ for \ce{Val_{10}-Lys^+-Val_{10}} 
(Table~\ref{table:peg-pept}). 
The protonated peptides were modeled with CHARMM forcefield. 
The water molecules were modeled with the modified-TIP3P (mTIP3P) model. 
The Rayleigh limit of the droplet were calculated with the surface tension of the
water model used at the simulation temperature \cite{vega2007,chen2007}.
At $T = 300$ K, the value of surface tension is
taken to be 0.0523 N/m and at $T = 350$~K to be 0.0432 N/m.

Newton's equations of motion were integrated using the velocity-Verlet algorithm 
with a time step of 1.0 fs.
All long range forces were computed directly.

For the peptides, two types of simulations were performed with NAMD v.2.12: 
(a) Equilibrium simulations where the droplet
was placed in a spherical cavity of radius 20.0 nm by using spherical boundary conditions. 
The cavity was sufficiently large to accommodate the shape fluctuations of the droplet. 
In the equilibrium simulations, the droplet reached vapor pressure equilibrium.
(b) Non-equilibrium simulations where the droplet was placed in vacuo and was evaporated.
We performed equilibrium simulations for each of the sequences at 300~K and 350~K, 
and non-equilibrium simulations at 350~K. 
The equilibrium runs were performed once at each temperature and the non-equilibrium runs in quintuples.
For PEG54 only equilibrium simulations were performed in the same manner as the
equilibrium simulations were performed for the protonated peptides. 
The number of water molecules ($\rm{N_{H_2O}}$) reported 
in Table~\ref{table:peg-pept} is the initial number in 
the droplet. In the equilibrium runs  $\approx 100$ \ce{H2O} molecules evaporated 
and created an equilibrium vapor pressure within the cavity. 

All the systems, with PEG and protonated peptides, were thermalized with the Langevin thermostat with the damping coefficient set to $1~\rm ps^{-1}$.

\begin{table}[htb!]
\caption{Systems and conditions of aqueous droplets with (a) PEG54 
and \ce{Na+} ions, and (b) protonated peptides and \ce{Na^+} ions.
In the first column, the system composition is shown.
The system composition is that for initializing equilibrium runs for the
protonated peptides and PEG54 and for performing evaporation
runs in vacuo (non-equilibrium setting) for the protonated peptides. 
The second column presents the initial
number of \ce{H2O} molecules in the droplet ($\rm{N_{H_2O}}$).
The third column shows the temperature ($T$) of the systems.
}
\begin{tabular}{|c|c|c|}
	\hline
Solutes                                   & $\rm{N_{H_2O}}$ & $T$ (K)  \\
\hline
PEG54 + 29 \ce{Na+}                       & $1\times 10^4$ & 300 \\
PEG54 + 25 \ce{Na+}                       & $1\times 10^4$ & 350 \\
PEG54 + 18 \ce{Na+}                       & $5\times 10^3$ & 300 \\
PEG54 + 18 \ce{Na+}                       & $5\times 10^3$ & 350 \\
	\hline
	\hline
\ce{(Val_{6}-Lys^+)_3-Val_{6} + 11Na^{+}}  & 2183  & 350 \\
\ce{(Val_{10}-Lys^+)_3-Val_{10} +11Na^{+}} & 2117  & 350 \\ 
\ce{(Val_{10}-Lys^+-Val_{10}) + 11Na^{+}}  & 2080  & 350 \\
	\hline
\end{tabular} 
\label{table:peg-pept}
\end{table}

\subsection{Myoglobin}

\begin{table*}[htb!]
\caption{Systems, conditions, and MD outcomes of protonated 
myoglobin\cite{watson1969} (1MBN and 1WLA) embedded in an aqueous droplet that also contains 
\ce{Na^+} ions.  
The first column shows the type of myoglobin and its charge state used in the various
MD runs.
In the second column $\rm{N_{H_2O}}$ has the same meaning as in Table~\ref{table:peg-pept}. 
In the third column, total charge of the droplet is the sum of the charge state of myoglobin
and the number of \ce{Na+} ions included in the droplet (details in the text).
In the fourth column, the initial conformation of the protonated myoglobin
to start the simulations is indicated. 
It is noted that the initial random coil conformation is generated from a zero 
charge bulk-phase equilibrium conformation for 1MBN, 
and from a gas-phase canonical charge conformation for 1WLA\cite{metwally2018chain}. 
More details on the preparation of the
initial conformation are found in the text.
``Fully equilibrated'' indicates that the charged protein is equilibrated within a droplet 
by restraining its position in the COM of the droplet or near its 
surface.  
In the sixth column, the outcome of MD simulations on protonated myoglobin's 
location is summarized. Length of simulation is indicated in
Fig.~\ref{fig:Xvstime}. In the seventh column the temperature, $T$, of the systems
is presented.}

\begin{tabular}{|c|c|c|c|c|c|c|}
	\hline
Type \&   & $\rm{N_{H_2O}}$       & Total     & Initial         & Initial location & MD outcome  & $T$ (K)  \\
charge     &                       & droplet &  myoglobin    & of charged    & of myoglobin's &      \\
state	   &                       & charge   & conformation         & myoglobin        & location &     \\

	\hline
$\rm 1MBN^{17+}$    &20058 & $+36$          & Random Coil  & Center           & Off-center           & 350 \\   
$\rm 1MBN^{23+}$    & 22494 & $+36$          & Random Coil  & Near surface     & Near surface        & 350  \\
$\rm 1MBN^{17+}$    & 22505 & $+40$          & Fully equilibrated  & Near surface  & Near surface    & 350  \\   
$\rm 1WLA^{22+}$[A] &22766  & $+47$          & Random Coil  & Center        & Partially extruded     & 370  \\   
                    &      &                &              &                  & \& rest dried-out    &  \\
$\rm 1WLA^{22+}$[A] &23179 & $+47$          & Random Coil  & Near surface     & Partially extruded   & 370    \\
                    &      &                &              &                  & \& rest dried-out    &  \\
$\rm 1WLA^{22+}$[C] &22401  & $+47$         & Fully extended & Center         & two sub-droplets       & 350  \\ 
                    &       &               &                &                & at protein's termini    &  \\
	\hline
\end{tabular} 
\label{table:myoglobin}
\end{table*}

We performed non-equilibrium simulations of charged aqueous droplets containing a charged myoglobin dimer (either 1MBN or 1WLA) and \ce{Na+} ions. The charge state of the proteins, the temperature and
the outcome of the simulations are shown in Table~\ref{table:myoglobin}.
1MBN is the sperm whale myoglobin\cite{watson1969}, obtained by Watson in 1969. 
1WLA is a wild-type recombinant horse heart myoglobin\cite{Marius1997}.  The sequence of aminoacids of the two myoglobin used in this study are similar, with 134 out of 153 identical amino acids. 
Both sequences were obtained with X-ray diffraction, 
but the resolution of 1MBN was 2.0~{\AA} and 1WLA 
was 1.7~{\AA}. 

It is known that non-equilibrium simulations are very sensitive  to the 
initial configuration of the system. 
Moreover, experimental techniques are not able to determine the exact conformation of the protein and its location within the droplet when the ESI sprayed droplet is at the nanometer scale. 
Therefore, we elected to start from multiple combinations of initial configurations where the
protein is found in a coiled state and an extended state. In addition, we performed simulations where the protein is initially located in the center  and near the droplet-vacuo interface.

\paragraph{1MBN} Simulations of \ce{1MBN^{17+}} and $\rm 1MBN^{23+}$ were performed using NAMD version 2.12. 
The charge pattern of the \ce{1MBN^{17+}} is as follows: All His residues are protonated (charge $+1$), 
all Asp and Glu residues are unprotonated (charge $-1$) except Asp27, Glu109 and Glu136 
which are protonated (charge 0). The charge pattern of the \ce{1MBN^{23+}} is as follows: 
All His residues are protonated (charge $+1$), all Asp and Glu residues are protonated (charge 0).

The protein was modeled with the CHARMM36m force field\cite{Huang2017} and the water was modeled with 
the mTIP3P model. The protein was solvated with $2.0 \times 10^4 - 2.25 \times 10^4$ \ce{H2O} 
molecules and the charge of the system was increased to $+36$ by adding \ce{Na+} ions
(13 \ce{Na+} ions in $\rm 1MBN^{23+}$ and 19 \ce{Na+} ions in \ce{1MBN^{17+}}) (Table~\ref{table:myoglobin}).
The initial configuration for 1MBN were generated by stretching the protein in gas phase, 
then setting the charge of every atom to zero and let the chain relax into a loosely 
coiled conformation. This yields an initial conformation that lacks a well-defined 
secondary structure. These coiled conformations were then solvated
and charged. We used the gaseous phase conformation in order to compare
with the simulations in Ref.~\cite{metwally2018chain}. In the Results and Discussion section
we comment on the disadvantages of
using a gaseous phase conformation within the droplet.

Additional simulations were performed where the COM of
\ce{1MBN^{17+}} and \ce{1MBN^{23+}} 
were restrained near 
the droplet surface (Table~\ref{table:myoglobin}) for relaxing the
protein conformation before the cavity was removed 
and evaporation in vacuo (without the presence of the
spherical cavity) was initiated. The relaxation of the conformation
lasted for 5~ns. The change in the radius of gyration and the root-mean-square relaxation
as a function of time are shown in Fig.~S1 in SI.

\paragraph{1WLA} For comparison, we used the charge patterns of 1WLA$^{22+}$[A] and 1WLA$^{22+}$[C] 
in Ref.~\cite{metwally2018chain}.
The exact charge pattern can be found in the Supporting Information of Ref.~\cite{metwally2018chain} 
\url{https://pubs.acs.org/doi/suppl/10.1021/acs.analchem.8b02926/suppl_file/ac8b02926_si_001.pdf}. 
The charge assignment of each residue and the simulation method is identical to that in 
Ref.~\cite{metwally2018chain}.

The protein was modeled with the CHARMM36 force field and the water was modeled with 
the TIP4P/2005 model\cite{Abascal2005}. The protein was solvated with $\approx 22500$ water molecules and the charge of the system was increased to $+47$ by adding \ce{Na+} ions (Table~\ref{table:myoglobin}. 
The Nos\`e-Hoover thermostat was used to thermalize the system at 370~K.
The droplet was placed in a cubic simulation box with sides of 999~nm. 
The cutoff distance were set to 333~nm and long range electrostatic forces were evaluate directly. 
Water and \ce{Na+} ions that traveled further than 100~nm from the COM of the system was removed every 250~ps. The trajectory stitching method was used to
continue the simulation once the \ce{H2O} molecules were removed. The length of the simulation was 40~ns.

The initial configuration for the 1WLA$^{22+}$ system were generated by relaxing the protein in gas phase for 100~ns. 
Specifically, it is written in Ref.~\cite{metwally2018chain}. ``Unfolded protein structures were initially produced by
heating aMb (1WLA without heme) from 320 to 450 K in vacuum using canonical charge states over 20 ns. 
Conformers generated toward the end of these runs served as starting points for droplet simulations.''
In Ref. ~\cite{metwally2018chain} no information was provided of how the initial conformation was prepared. 
By experimentation we found that 1WLA$^{22+}$[A], [C] in the gaseous phase (bare of
water) were extended when no cut-off was used in the electrostatic interactions, and it remained 
a loose coil when cut-off was used. Test simulations were performed with 1WLA$^{22+}$[C] 
in the extended form with no cut-offs. A fully extended conformation was obtained by 
simulating it in gas phase for ~100~ns. The fully extended conformation was then solvated 
with layers of \ce{H2O} molecules to form a cylindrically shaped droplet.
As the system relaxes, the droplet divided into two sub-droplets, each solvating the ends of the protein. 
We think that the origin of this separation 
is due to the interplay of two competing factors: minimization of the surface
area of a water droplet, and maximization of the protein solvation. 
In order to have a compact droplet, 
the coiled conformation was used as an initial conformation. 

It is emphasized here, 
that in the simulations of  \ce{1MBN^{17+}} the manner in which the initial coil conformation 
was prepared was different from that followed for 1WLA$^{22+}$[A] and [C].

\paragraph{Drawbacks in simulations of proteins in droplets} 
Here we note the uncertainties
in the modeling of 1WLA$^{22+}$[A] and [C] as presented in Ref.~\cite{metwally2018chain}.
TIP4P/2005 reproduces the surface tension\cite{vega2007,chen2007} of water over a range of temperature, 
however, the CHARMM force field for proteins is parametrized with respect to a 
modified version of TIP3P (mTIP3P). 
A charged protein in general might be in a different conformation when used with water models
that it has not been parametrized for it.
In our opinion, it is more important to model well the interaction between the water molecules and the protein side chains than the surface tension of the droplet. 
Although mTIP3P reproduces lower surface tension of water than TIP4P/2005, the surface tension is  
a parameter in the fissility parameter (Eq.~\ref{eq:fissility}) that will
determine the amount of charge that the system can hold for the particular surface tension. 
Therefore, using a water model that replicates the surface tension of water may not be necessary. 

The trajectory stitching method utilizes a pseudo-PBC approach, where the droplet 
is placed in a very large simulation box and where long range electrostatic 
interactions are calculated directly as opposed to using conventional Ewald sum methods. 
The evaporated water molecules are removed periodically and the velocity of each molecule 
is reassigned according to the Boltzmann distribution after evaporated water have been removed. 
However, under non-equilibrium conditions where rapid solvent evaporation
takes place, the Boltzmann distribution may not be followed.
Moreover,
the simulations have been performed in very elevated temperature (370~K) that
may not be relevant to experimental conditions\cite{soleilhacantoine2015, gibsoncook2013}.
Methods for evaporating \ce{H2O} molecules without relying on
the trajectory stitching method and for efficient evaporation
of the solvent molecules even at room temperature have been proposed\cite{consta2022atomistic}. 
The selection of the initial conformation of the protein from the gaseous state
is another factor that biases the outcome of the MD trajectories. The effect of this
bias is discussed in the Results and Discussion.

\section{Results and Discussion}

\subsection{Charging of PEG provides insight beyond
minute nanodroplets}

\begin{figure}[htbp]
  \begin{subfigure}{.49\textwidth}
        \centering
        \includegraphics[width=\linewidth]{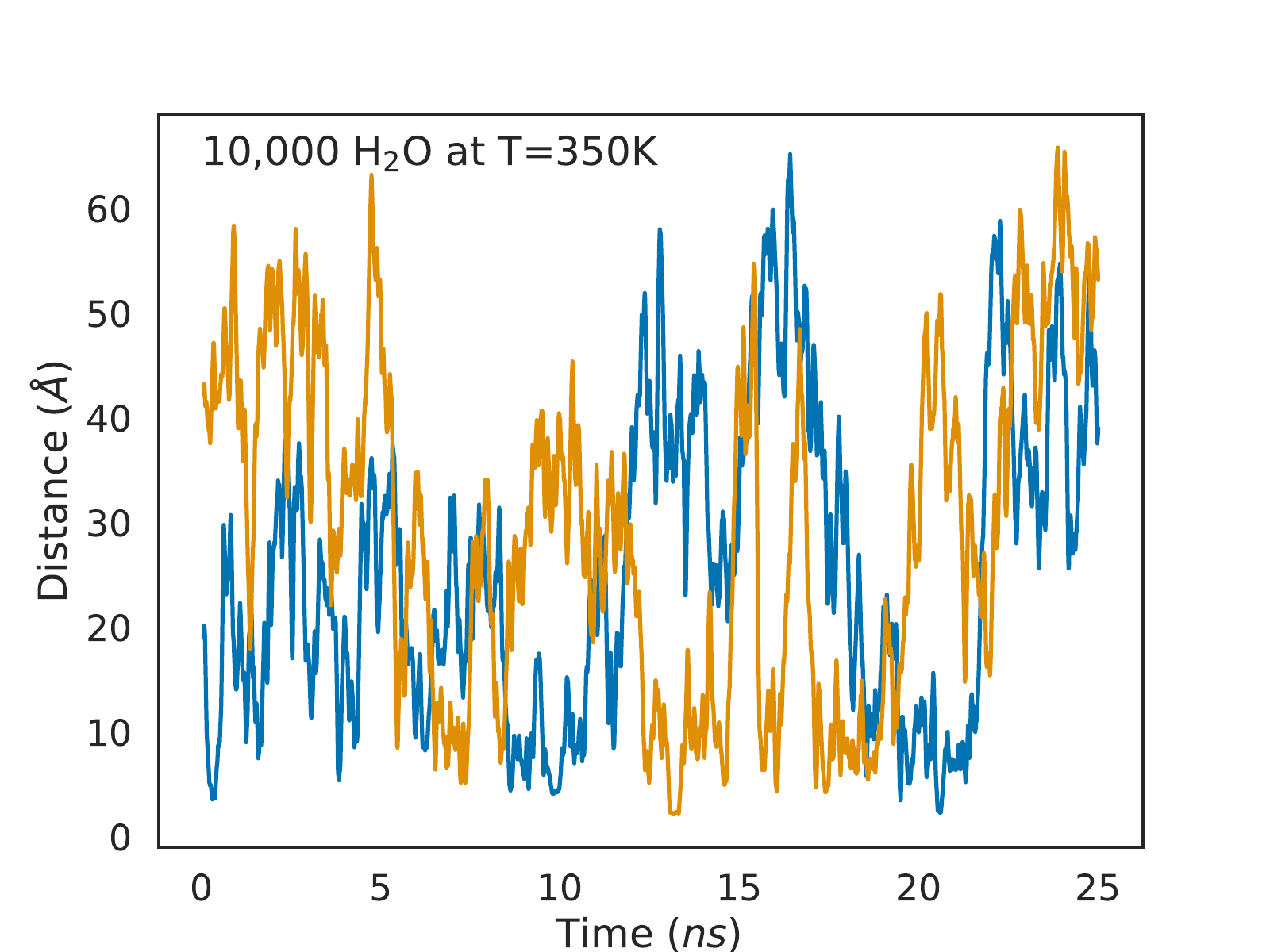}
        \caption{ }
    \end{subfigure} %
  \begin{subfigure}{.49\textwidth}
        \centering
        \includegraphics[width=\linewidth]{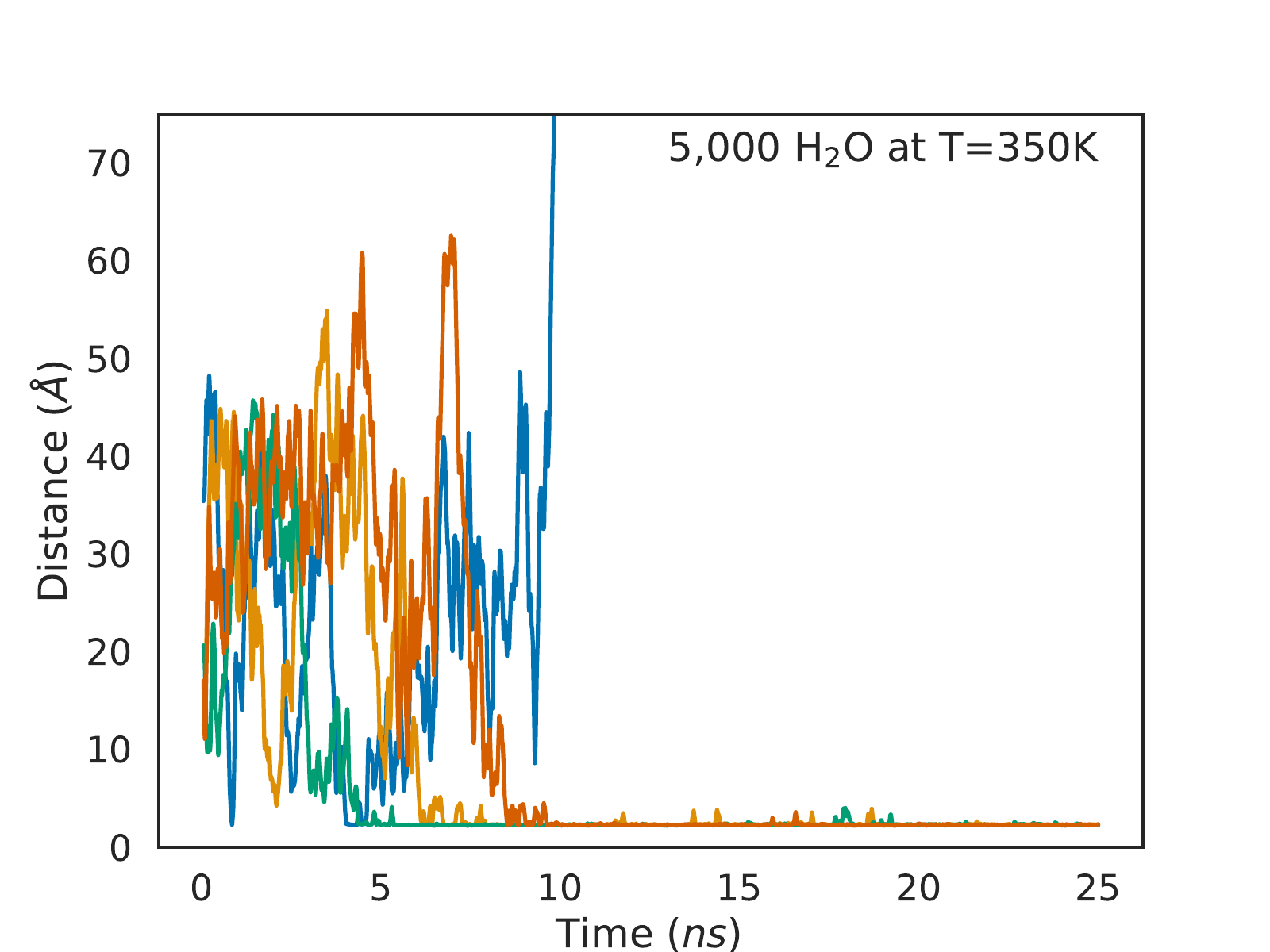}
        \caption{ }
    \end{subfigure} %
    \caption{Typical sodiation events of PEG in aqueous droplets as a function of time.
       (a) $10 \times 10^3$ \ce{H2O} molecules, 25 \ce{Na+} ions and PEG54
       and (b) $5 \times 10^3$ \ce{H2O} molecules, 18 \ce{Na+} ions and PEG54, 
       at 350~K (see Table~\ref{table:peg-pept}). 
    The y-axis measures the least distance of a
    particular \ce{Na+} ion from any PEG54 oxygen atom.
    Several transient sodiation events were
    observed in the course of the simulations, but for clarity, 
    only the PEG54 binding to two and four \ce{Na+} ions  
    in systems (a) and (b), respectively, is shown.
    Similar graphs but at 300~K are shown in Fig.~S2 in SI.
  }
    \label{fig:PEG54-10K-5K-combined}
\end{figure}

\begin{figure}[htbp]
  
        \centering
       \includegraphics[width=\linewidth,trim={0 0 0 0},clip]{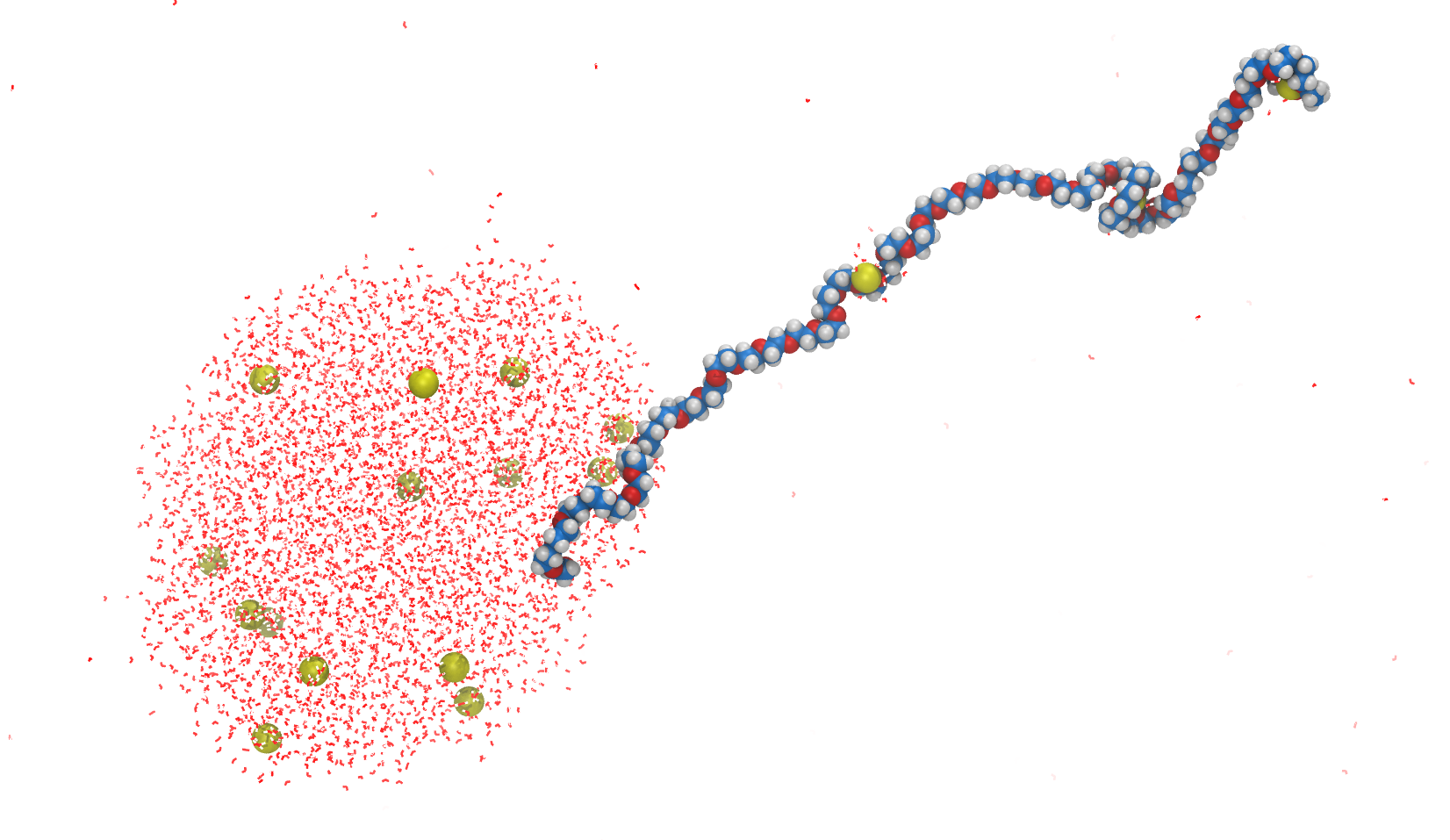}
    
	\caption{Typical snapshot of PEG54 at 350~K in a droplet comprised $5\times 10^3$ \ce{H2O} molecules (O colored in red and H in white) and 18 \ce{Na+} (colored in yellow) ions (see Table~\ref{table:peg-pept}).
	PEG54 captures three \ce{Na+} ions and escapes. The sizes of PEG and \ce{Na+} ions have been enlarged
	relative to \ce{H2O} molecules for visualization purposes.}
    \label{fig:PEG54-snapshot}
\end{figure}

\begin{figure}[htbp]
  \begin{subfigure}{.49\textwidth}
        \centering
        \includegraphics[width=\linewidth]{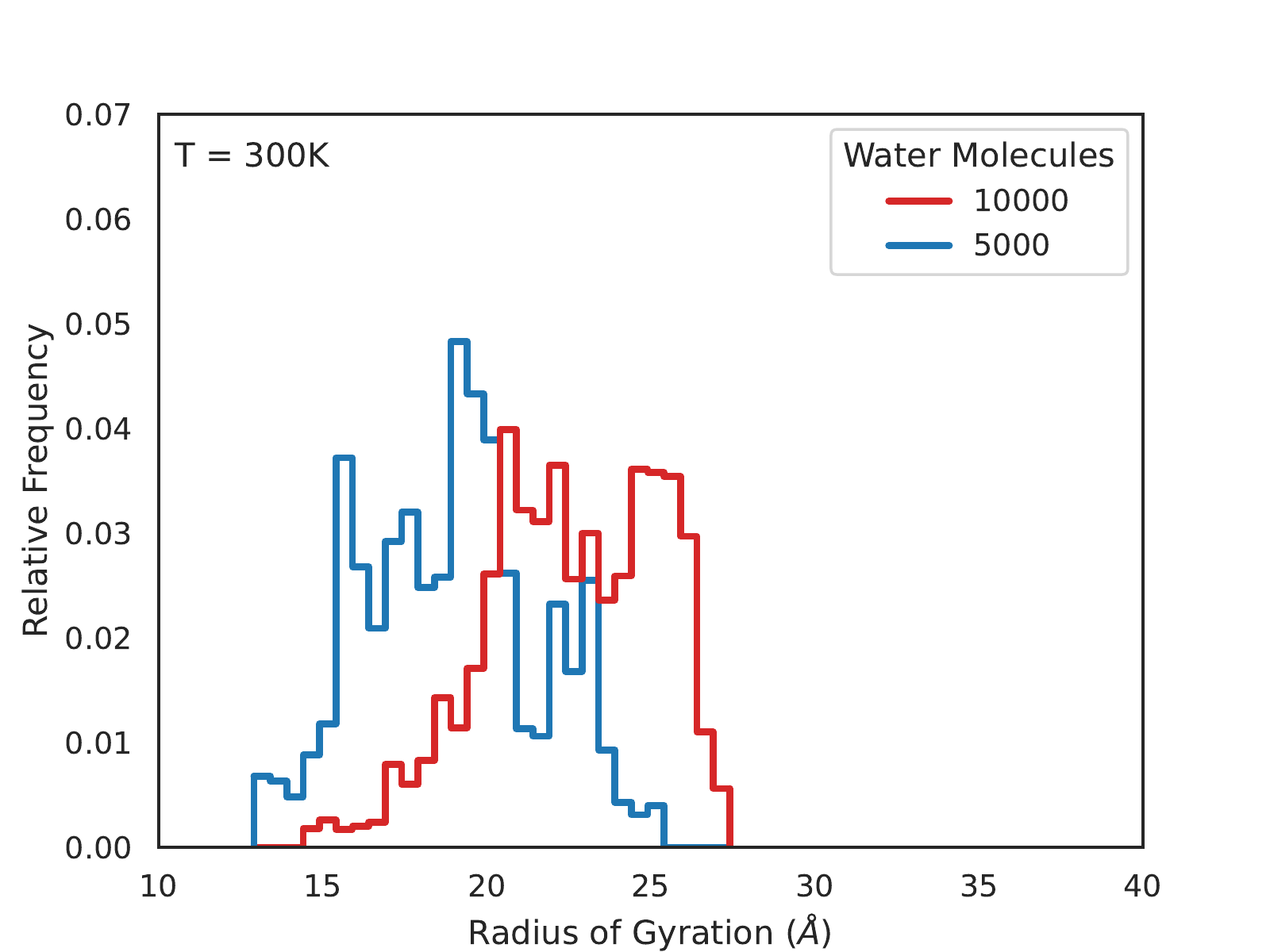}
        \caption{ }
    \end{subfigure} %
  \begin{subfigure}{.49\textwidth}
        \centering
        \includegraphics[width=\linewidth]{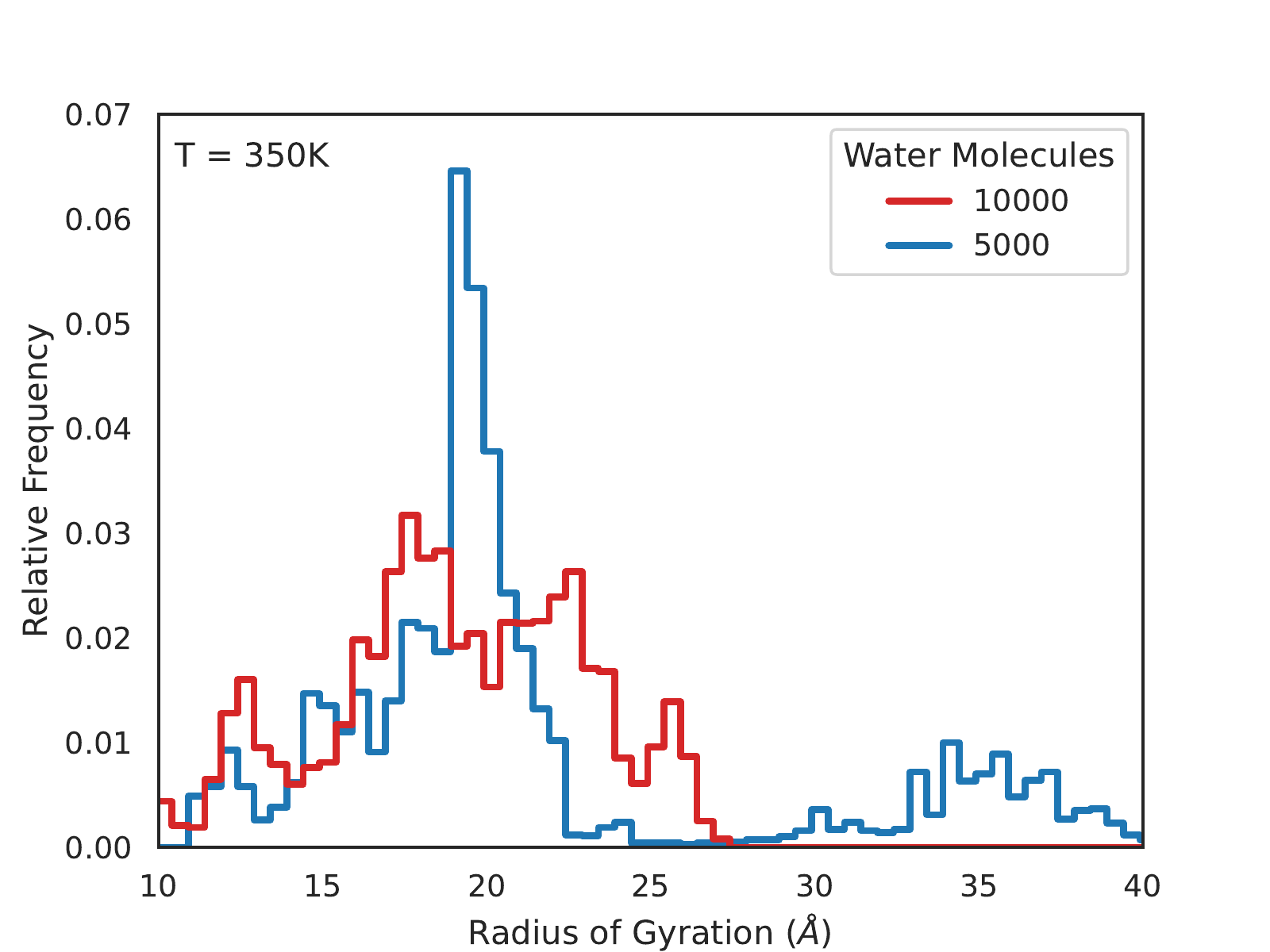}
        \caption{ }
    \end{subfigure} %
	\caption{Distribution of radius of gyration of PEG54 at (a) $T=300$~K and (b) $T=350$~K 
	  in droplets comprised  $5 \times 10^3$ \ce{H2O} molecules and 18 \ce{Na+} ions,
	  and $10\times 10^3$ \ce{H2O} molecules and 25-29 \ce{Na+} ions (Table~\ref{table:peg-pept})
	  The time evolution of RG is shown in Fig.~S3 in SI. 
}
    \label{fig:PEG54-rgyr-dist}
\end{figure}

PEG may be considered to share some commonalities with
intrinsically disordered proteins (IDPs) because of 
their flexibility in their backbone.
PEG has the great advantage that it allows one to directly observe 
the transfer of ions from \ce{H2O} to the chain and its extrusion from a charged droplet\cite{constapeg1}. 
Even though we have studied the PEG charging mechanism 
\cite{constapeg1, constapeg2, constapeg2015, oh2017charging, InOhPHDthesis, sharawy2014effect}
a remaining question is whether our observations in small nanoscopic systems 
are transferable to larger droplet
sizes and longer macromolecules than those atomistically modeled.
To mimic these large droplets,
we scale down the problem to a large droplet size relative to the length of PEG. 
Specifically, we model PEG54 in droplets comprised $10 \times 10^3$ \ce{H2O} molecules
and $5 \times 10^3$ \ce{H2O} molecules and \ce{Na+} ions at $T=300$~K and 350~K 
(Table~\ref{table:peg-pept}). 
During the entire trajectory within the spherical cavity a few \ce{Na+} ions
escape from the parent droplet. The droplet composed initially of $5 \times 10^3$ \ce{H2O} molecules  
emits one \ce{Na+} and that of $10 \times 10^3$ \ce{H2O} molecules emits three \ce{Na+} ions. 
Decisive factors in the charging and release of PEG and by extension in 
any linear macromolecule are: the ratio of the charge on the macroion while still attached
in the droplet's body
to the free charge in the 
droplet\cite{constapeg1, constapeg2, oh2017charging, InOhPHDthesis}, the temperature of the system,
and the presence of  counterions\cite{sharawy2014effect}.  
Here we identify a new factor, which is the effect of the droplet curvature on the PEG
conformation and thus, on the charge state.

Figure~\ref{fig:PEG54-10K-5K-combined}~(a) and (b) show the extent of time 
that \ce{Na+} ions bind to the
PEG backbone. In droplets comprised $10 \times 10^3$ \ce{H2O} molecules at both 300~K and 350~K,
\ce{Na+} ions transiently bind to PEG. Differently, in a smaller droplet, 
of $5 \times 10^3$ \ce{H2O} molecules \ce{Na+} ions bind long enough (permanent sodiation) for 
the sodiated PEG to extrude from the droplet as shown in Fig.~\ref{fig:PEG54-snapshot}.  
In both droplet sizes, \ce{Na+} are present for binding. The difficulty of
a long-living sodiation in a larger droplet comes from the PEG conformation.

Figure~\ref{fig:PEG54-rgyr-dist}~(a) and (b) show the histograms of PEG54 radius of gyration (RG) for 
two droplet sizes. In both temperatures the conformation of PEG in the $10 \times 10^3$ \ce{H2O}-molecule
droplet is more extended than in the $5 \times 10^3$ \ce{H2O}-molecule 
droplet. In the $5 \times 10^3$ \ce{H2O}-molecule
droplet, at $T=350$~K (Fig.~\ref{fig:PEG54-rgyr-dist}~(b)) there is a second broad peak between 30~{\AA}-40~{\AA}, which is due to the
extruded linear sodiated PEG54. The more extended PEG54 conformation in the larger
droplet is caused by the droplet curvature.
Thus, even though \ce{Na+} are available for binding, the droplet curvature prevents the coiling
of the chain that may lead to a long-living sodiation.

Here we synthesize the overall picture that we have obtained by systematic simulations of charging of 
PEG with \ce{Na+} (also \ce{Li+} and \ce{Ca^{2+}} ions) in previous 
research\cite{constapeg1, constapeg2, constapeg2015, oh2017charging, InOhPHDthesis} 
and in the present studies 
in aqueous nanodroplets composed 
up to ten thousand \ce{H2O} molecules (equimolar radius 4.2~nm). 
The new results allow us to understand the mechanism in droplet sizes 
that are not accessible to atomistic modeling yet.

The general mechanism we have identified for the charging of
PEG regardless of its length and the droplet's size is as follows:
Evidence from previous
simulations\cite{constapeg1, constapeg2, constapeg2015} has shown that 
a PEG of specific length can be charged in droplets of various sizes and partially extrude.
Specifically, in previous research\cite{constapeg2015} 
we found that PEG64 can be charged via long-living (we call it permanent) sodiation 
in droplets with $\approx 7000$, $3500$ and $2000$ \ce{H2O} 
molecules. Previous results are complemented by the present study that shows for the
first time, that
for a PEG of specific length there is a \textit{critical} droplet size 
beyond which, PEG is charged intermittently. The life-time of the sodiation
events is shorter than the time required for the PEG backbone to unzip from
the droplet surface and release by carrying the ion with it.

We have found that the first permanent sodiation takes place on the droplet surface in 10\%-15\% overall 
charge
below the RL in droplets of different sizes\cite{constapeg1, constapeg2, constapeg2015}. 
If we compare this value with the $X$ (defined in Eq.~1) values that a Rayleigh jet 
is formed\cite{consta2022atomistic}, 
we conclude
that it is the range of 10\% - 15\% charge below the RL where the fluctuations will be intensified 
and the critical point of the Rayleigh instability is approached. Thus, the first transfer
of an ion to PEG
occurs at the onset of the intense shape fluctuations near the RL. 
The captured \ce{Na+} ions may be escaping ions from the water body due to intended 
formation of a jet. Since PEG may capture the ions, the formation of the jet may be
avoided in certain cases. We expect that an entire PEG or a segment of it can be also captured 
by a Rayleigh jet when formed. For droplets with fewer than $\approx 1000$ ~\ce{H2O} there are some
more details that enter the charging mechanism because of their large
relative shape fluctuations that bring the value of $X$ much less than 1 for 
droplet stability\cite{kwan2020bridging}.

Once an ion is ``permanently'' captured by PEG, a part of the PEG chain extrudes.
We have found that in the majority of the PEG simulations at 300~K and 350~K
the chain partially extrudes and then a part of it dries-out.
Drying-out has been observed significantly more frequently than ejection for
the following reason: There are many more free charges (\ce{Na+}, \ce{H3O+}) in 
a droplet than macroions. The likelihood of the single ion release is
much higher than of the macroion's. Since single solvated ions are released, and
also the chain immersed in the droplet may be charged, the droplet may not have
a critical charge to repel a segment of the chain. 

Now, we ask the question: In which droplet size does PEG sodiation occur?
The present in combination with previous studies\cite{constapeg1, constapeg2, constapeg2015, oh2017charging, InOhPHDthesis, sharawy2014effect} show that charging of PEG
is highly likely to occur in droplets much larger than the ones
that can be atomistically simulated. 
The longer the PEG, the larger the droplets in which it can be
charged. There is a \textit{critical} droplet size above which short-lived charging 
of PEG may occur. Lower charge
density in larger droplets may be an additional factor
that prevents the ejection of a  macroion from the droplet.

Similar mechanisms need to be considered for the charging of proteins.
Unfortunately, the proteins are very restrictive to a systematic study of their
charging and possible extrusion mechanism from droplets. Moreover, the
outcomes of simulations may be strongly biased by the selection of the protein's initial conformation.
A similar bias does not hold for at least the shorter chains of PEG.

If we were to transfer the transient sodiation to the protonation of a protein
we may infer that changes in the solvation of
a protein due to droplet size and the time required for conformational
changes will affect a protein's ability to capture a proton.
Thus, high proton availability and an extended protein conformation
does not necessarily lead to a higher protein charge state relative
to a less extended conformation, if considerable conformational changes
are required for the extended conformation to host the proton.

\subsection{Release of protonated Valine-Lysine peptides from nanodroplets - A rare event}

\begin{figure}[htbp]
   \centering
    \begin{subfigure}{.49\textwidth}
        \centering
        \includegraphics[width=\linewidth]{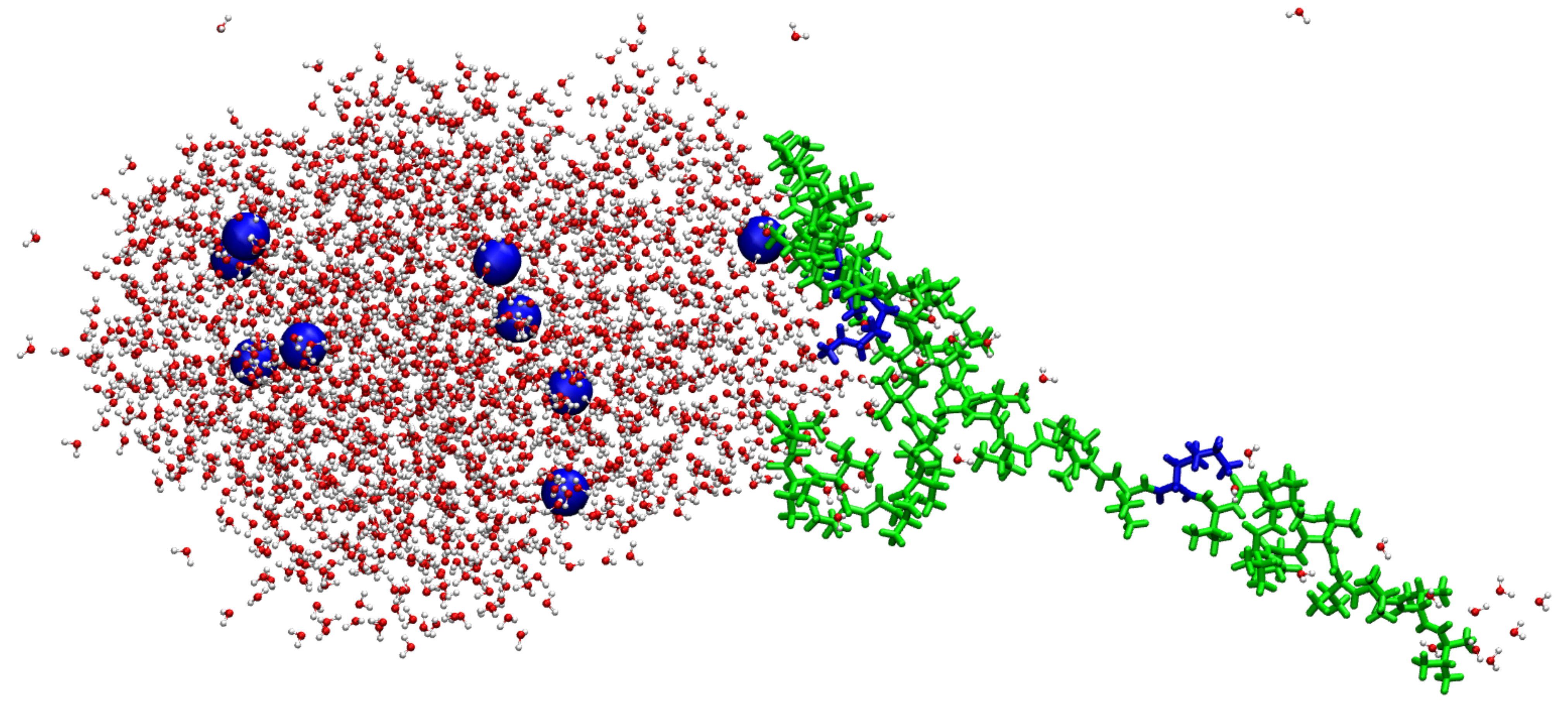}
        \caption{}
    \end{subfigure} %
    \begin{subfigure}{.49\textwidth}
        \centering
        \includegraphics[width=\linewidth]{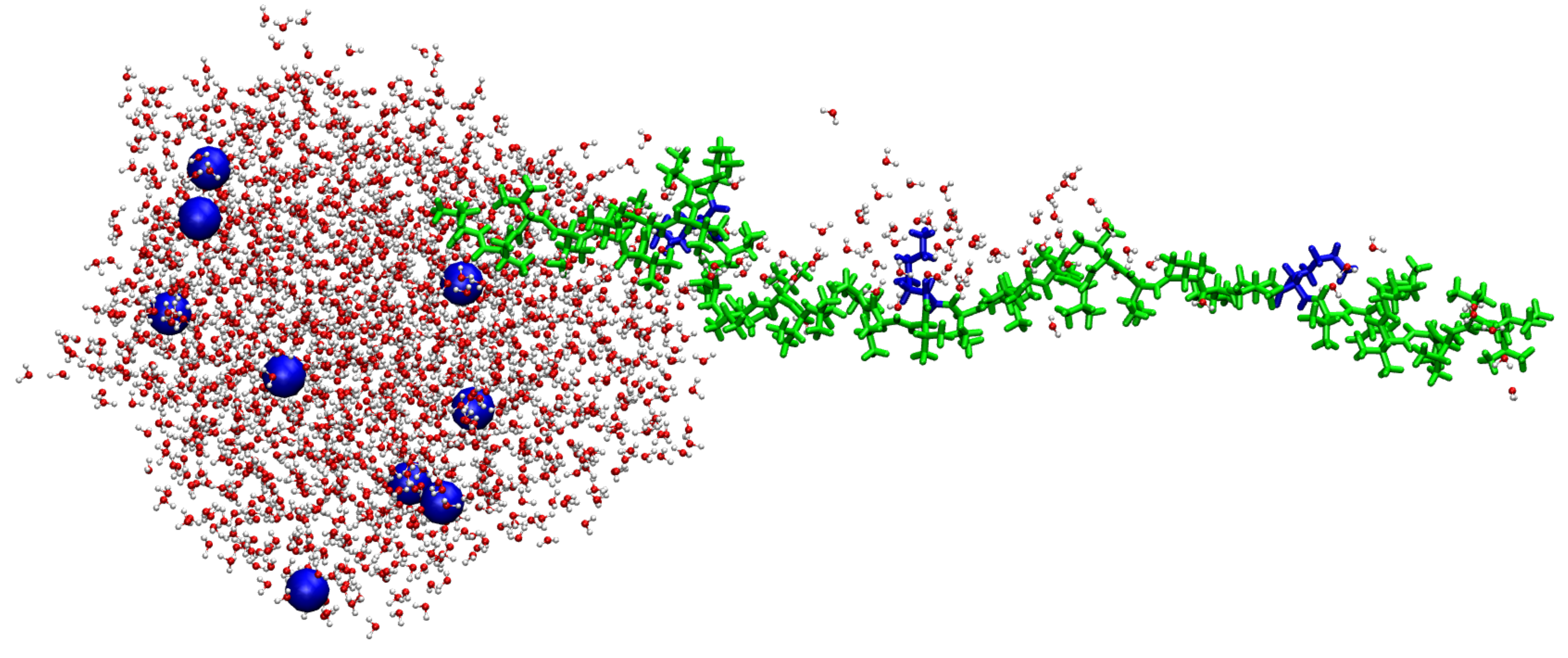}
        \caption{}
    \end{subfigure} %
    \begin{subfigure}{.49\textwidth}
        \centering
        \includegraphics[width=\linewidth]{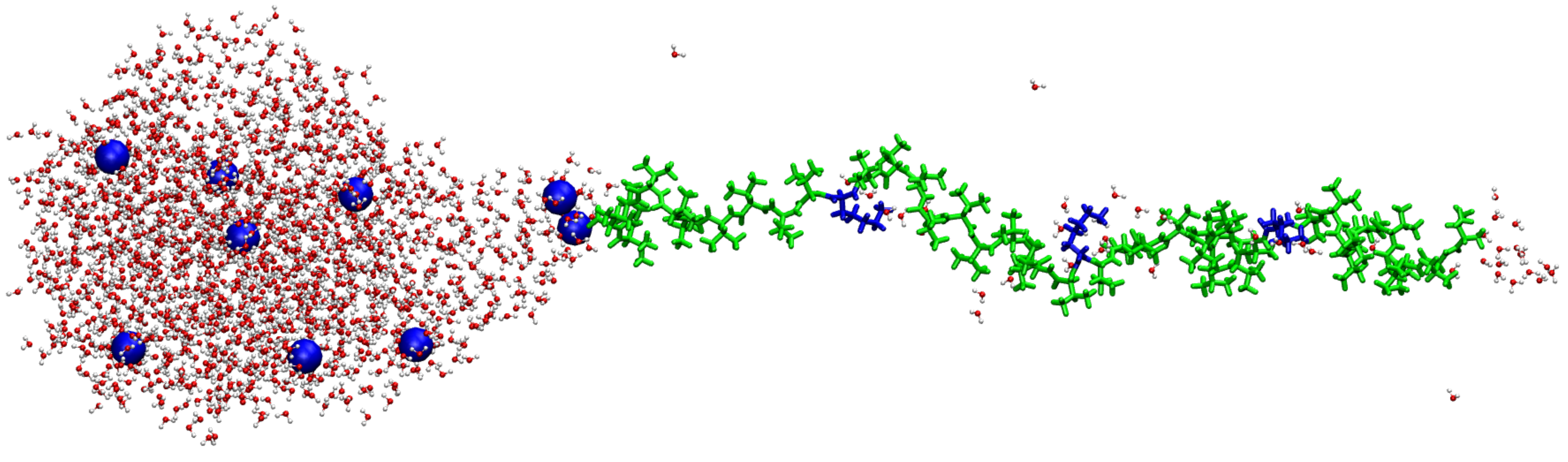}
        \caption{}
    \end{subfigure} %
  \caption{\label{fig:val10lysx3_ext}
  Typical snapshots of the \textit{rare} event of \ce{[(Val_{10}-Lys^+)_3-Val_{10}]} ejection
  from an aqueous droplet.
  The lysine and valine residues are shown along the chain by the blue and green colour
  residues, respectively,
  and the \ce{Na^+} ions are represented by blue spheres. 
  The O sites of the \ce{H2O} are shown by red spheres and the H sites by white.
  (a) First step of the extrusion of a solvated protonated peptide from a droplet composed
  of $\approx 1925$ \ce{H2O} molecules. We consider this configuration 
  as the time origin. (b) Further extrusion of the chain from a droplet composed
  of $\approx 1735$ \ce{H2O} molecules at 4.4 ns from the time 0. (c) Complete ejection
  of the chain from a droplet composed of $\approx 1694$ \ce{H2O} molecules
  at 5.2 ns from time 0. The ejection is
  assisted by two \ce{Na+} ions at its attachment point to 
  the droplet. 
  Additional snapshots are shown in Fig.~S3 in SI. 
}

\end{figure}

\begin{figure}[htbp]
    \begin{subfigure}{.60\textwidth}
        \centering
        \includegraphics[width=.60\linewidth]{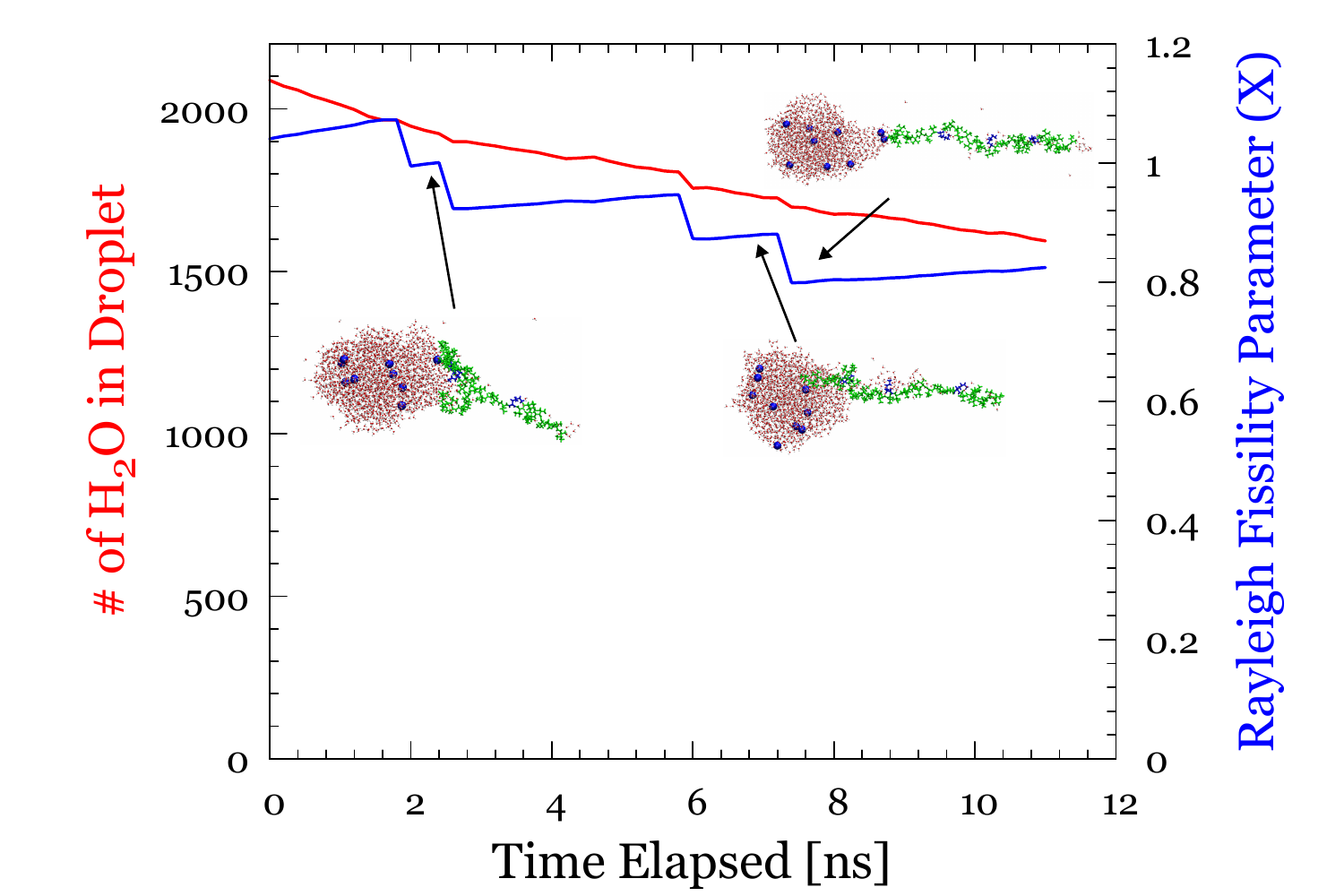}
        \caption{}
    \end{subfigure} %
    \begin{subfigure}{.60\textwidth}
        \centering
        \includegraphics[width=.60\linewidth]{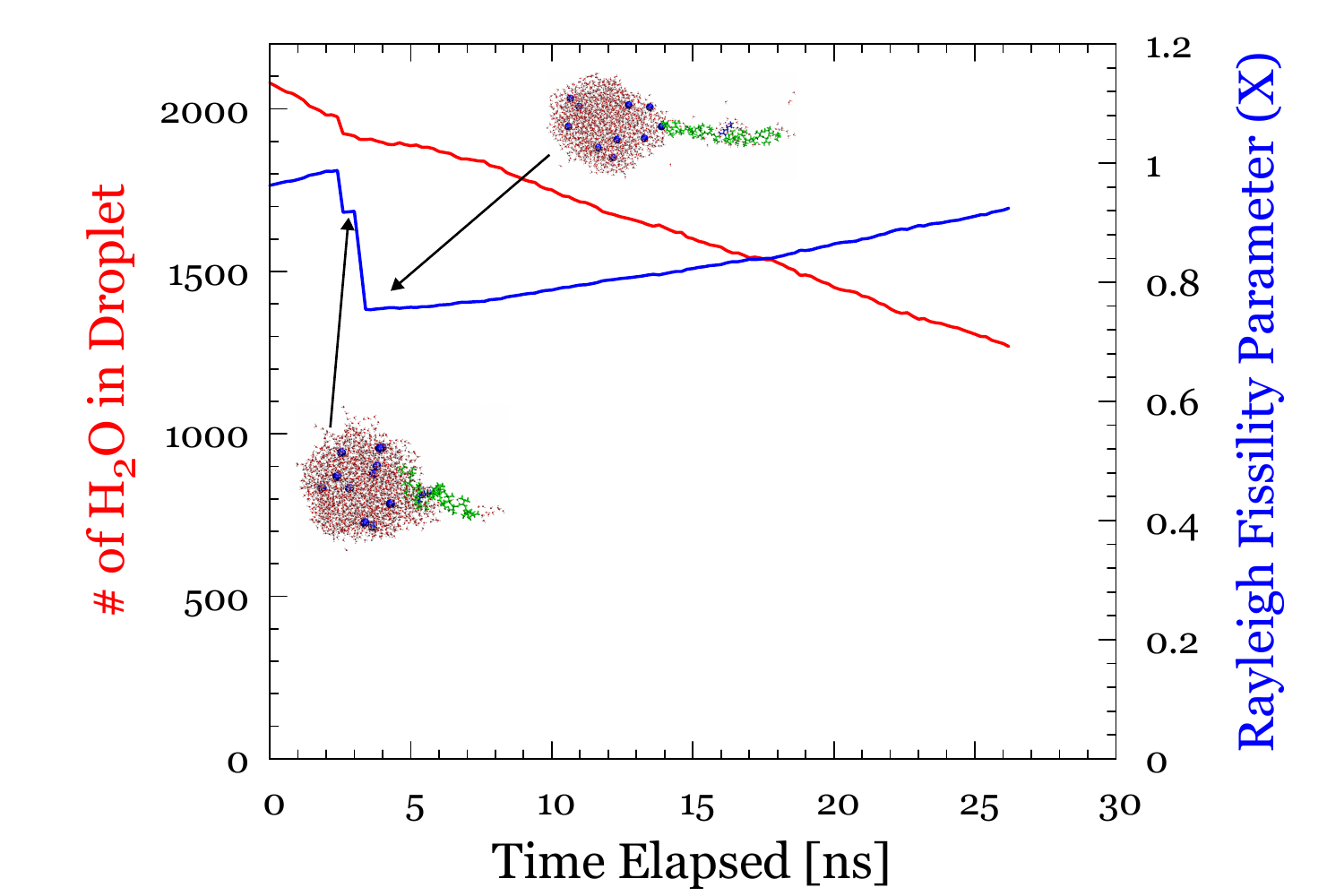}
        \caption{}
    \end{subfigure} %
    \begin{subfigure}{.60\textwidth}
        \centering
        \includegraphics[width=.60\linewidth]{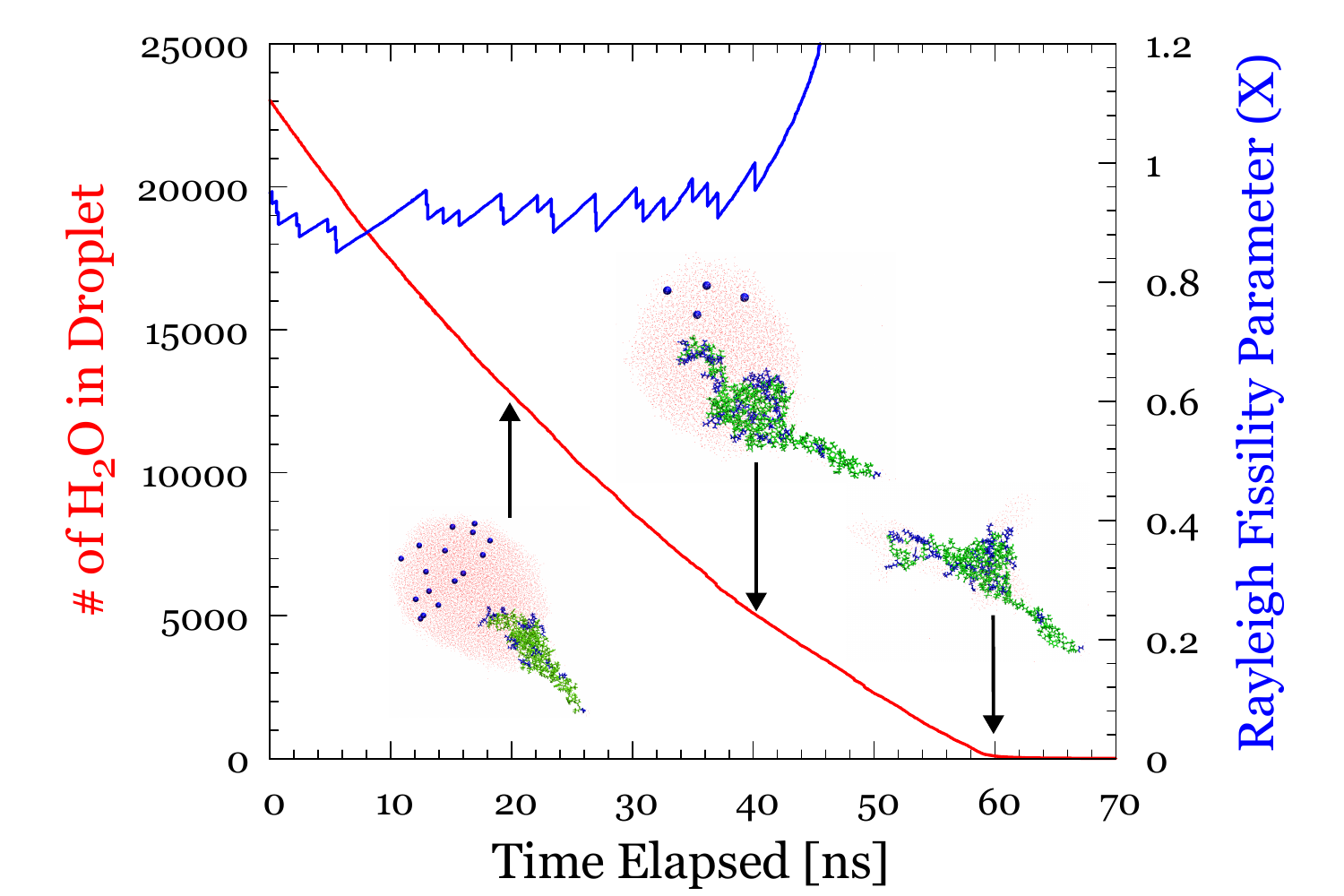}
        \caption{}
    \end{subfigure} %
       \caption{Time evolution 
	 of number of \ce{H2O} attached to the droplet and Rayleigh fissility parameter, $X$ defined
	 in Eq.~\ref{eq:fissility} in a drying-out process.
	 (a) \ce{(Val_{10}Lys+)_3Val_{10}}. (b) \ce{Val_{10}Lys+Val_{10}}. (c) 1WLA$^{22+}$. 
       $X$ is calculated by assuming a spherical shape of the droplet. 
       It is noted than in (c) $X > 1$ beyond 
       40~ns because the protein dries-out.
     In the estimation of $X$ only the charges within the droplet are included.} 
\label{fig:Xvstime}
\end{figure}

Droplets comprised
$\approx 2100$ \ce{H2O} molecules, a peptide comprised Lys+ and Val residues (Table~\ref{table:peg-pept})
and \ce{Na+} ions were simulated under equilibrium conditions (where an equilibrium is established between the droplet and its
vapor within a spherical cavity) and non-equilibrium 
evaporation runs in vacuo. The RL for these droplets at their initial size
is $+13$ if we consider that the peptide is
equivalent to 215~\ce{H2O} molecules. Droplets that initially have charge $+14$, release 
one or two \ce{Na+} ions.
Direct visualization of the systems shows that the Val residues lie on the surface of
the droplet, and the side chains of the protonated lysine residue point toward the interior.

It is noted that, in a nanometer scale droplet, hydrophobic and hydrophilic residues behave opposite to that of bulk solution. 
In a bulk solution, the hydrophobic residues prefer to aggregate and form a hydrophobic core, 
buried in the interior of a protein while the charged (hydrophilic) residues lie 
on the exterior of the protein. 
When a quasi-linear and flexible protein is located near the surface of a nanoscale droplet, 
the hydrophobic core may unfold and lie on the surface of the droplet, 
solvated by minimal amount of water in order to lower the solvation energy penalty. 
The hydrophilic residue, in general prefer to be well solvated, 
therefore when they are forced to the surface, the side group will tend to point inwards, 
which enable solvation stabilization. 
This is also supported by experimental data in electron cryo-microscopy where it 
has been shown that proteins tend to adsorb to the air–water interface with a preferred orientation, 
or that they may even become 
partially or fully unfolded at the interface \cite{Glaeser2018}.

Equilibrium simulations (where vapor and droplet are at equilibrium
within a spherical cavity) are used to establish a reference point for 
comparisons of the peptide's conformations. Initially, 
in the evaporation simulations (non-equilibrium) in
vacuo where the peptide does not extrude, 
the protonated peptides have the same conformation as in the equilibrium 
simulations. Moreover, no extrusion was found in the equilibrium runs.

In the non-equilibrium simulations it is found that in all cases the extrusion of the chain occur 
less frequently than the drying out. For \ce{(Val_{6}-Lys^+)_3-Val_{6}}, 
in all five evaporation simulations performed, protonated peptide extrusion was not observed, 
instead the droplet dries out.
During the course of the simulation, \ce{Na^+} can be seen ejecting from the droplet,
which indicates, that the ions are much more mobile than a chain-like macromolecule.

We increased the number of valine residues in between lysine residues,
to investigate whether the valine residue compact similar to
what we observed with pure poly-valine - and whether a
hydrophobic core might form in this case.
For \ce{[(Val_{10}-Lys^+)_3-Val_{10}]} three out of the five runs did not show any extrusion, 
one run showed partial extrusion and one run ejection of the protonated peptide. 
Typical snapshots from the ejection process are shown in Fig.~\ref{fig:val10lysx3_ext}.
It is noted that the ejection was assisted by the presence of two \ce{Na+} in a conical fluctuation
at the end of the chain 
found in contact with the droplet (Fig.~\ref{fig:val10lysx3_ext}~(c)). 

For \ce{[Val_{10}-Lys^+-Val_{10}]} 
extrusion was observed only in one case out of five runs. 
Similarly to \ce{[(Val_{10}-Lys^+)_3-Val_{10}]}, the extrusion was also assisted by 
the \ce{Na+} ions accumulated in a conical shape fluctuation. 
Typical snapshots of this system are shown in Fig.~S5 in SI.

Figure~\ref{fig:Xvstime}~(a) and (b) show the reduction of the \ce{H2O} molecules
and the variations in the value of $X$ (see Eq.~\ref{eq:fissility}) 
for \ce{(Val_{10}Lys+)_3Val_{10}} and \ce{Val_{10}Lys+Val_{10}}, respectively,
as a function of time in a drying-out process. 
Large decrease in the value of $X$ is associated with the release of a macroion
segment as shown in the snapshots included in the figure.

From all the simulations that were performed, even with highly hydrophobic peptides, it is 
  found that the partial extrusion is of low frequency
and the ejection of a protonated peptide is even more rare. 
This is due to the fact that the \ce{Na+} ions may
be released faster than the chain, and that even the hydrophobic residues 
form H-B with the droplet\cite{InOhPHDthesis}. 
The interplay between solvation of the chain and repulsion from the charges
of the same sign in the droplet has been discussed in detail 
in previous articles\cite{consta2012, constacpl2016}. It is noted here that
the charges of the same sign within the droplet 
can be the part of the solvated chain itself that may be charged.
Thus, the presence of free ions is not the only source for repulsion, and in fact,
the free ions are not necessary for triggering a chain extrusion.

In previous work\cite{kwan2021relation, consta2022atomistic} 
we demonstrated the significance of the conical 
fluctuations on
the droplet surface in the release of simple ions and in the
formulation of the ion evaporation mechanism\cite{iribarne1976, iribarne1979}. 
It appears that a rare event, where a conical fluctuation captures a protein's charged site
or it allows for other ions (e.g. \ce{H3O+} or \ce{Na+}) to assist in the repulsion of the chain, 
may lead to partial chain extrusion or detachment from the droplet. 

For the conical fluctuations to play a role in the extrusion of the chain, 
the droplet should be near the RL.
The fact that an extrusion may take place near the 
RL is along J. Fenn's conjecture that the charge
density should be appropriate for macorion desorption\cite{fenn1993ion, fenn1997electrospray}. 
However, Fenn's conjecture did not
consider the critical role of the conical fluctuations in the ejection mechanisms. 
For PEG a very similar
mechanism of extrusion has been found\cite{constapeg1, constapeg2}. 

\subsection{Charged myoglobin in aqueous droplets - A Taylor cone formation}

\paragraph{1MBN$^{23+}$} A semi-compact configuration of 1MBN$^{23+}$ was generated by 
removing all partial charges on the protein, and letting it equilibrate for 1~ns. 
We then solvated the protein in a droplet of 
$\sim 22500$ \ce{H2O} molecules, restored its charge, and 
added 13 $\rm{Na^+}$ to bring the droplet's overall charge slightly below the RL (Table~\ref{table:myoglobin}).
The RL for this droplet size corresponds to charge $+43.8$.
The protein was placed both in the center of the droplet
and near the surface. 
We have only observed evaporation of water, ejection of simple ions and protein undergoing conformational changes. 
In both runs, the protein was well solvated with water even after 25~ns, thus no partial extrusion
of the protein was observed.

\begin{figure}[htbp]
        \centering
        \includegraphics[width=\linewidth]{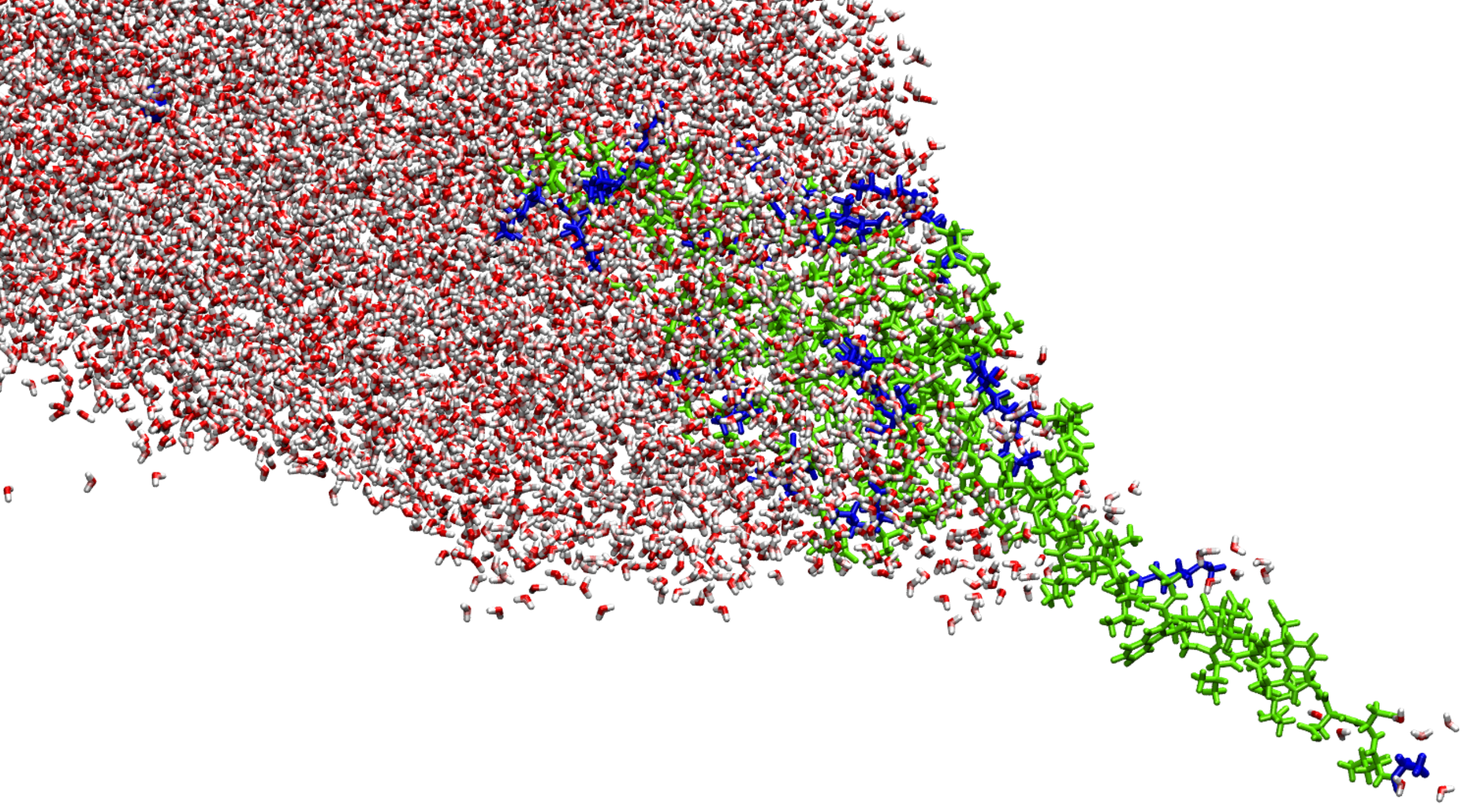}
        \caption{Magnification of a segment of a droplet to show 
	  the partial extrusion of \ce{1WLA^{22+}[A]} from
	an aqueous droplet (see Table~\ref{table:myoglobin}). The charged residues are colored in blue and the uncharged residues are colored in green. $\rm Na^+$ ions are depicted by blue spheres. The color coding
	of the \ce{H2O} molecules is the same as in Fig.~\ref{fig:PEG54-snapshot}.}
\label{fig:extrusion}
\end{figure}

\paragraph{Coiled 1WLA$^{22+}$}
A coiled 1WLA equilibrated in the gaseous phase was placed in the droplet's center and its charge
was restored to $+22$ (details are described in the ``Systems and Simulation Methods'' Section). 
It was observed that almost all \ce{Na+} ions diffused near the surface in 1~ns 
while the protein diffused much slower toward the surface.
The protein is expected to migrate near the surface because of the conducting nature
of the droplet. 
The diffusion time of 1WLA$^{22+}$ from the center to the surface was at 
least twice as long as those reported in Ref.~\cite{metwally2018chain}. This difference may not be 
critical for the protein extrusion.
The droplet shape underwent a significant fluctuation when the droplet moved near the surface. 
Snapshots of the entire process up to the protein's drying is shown in Fig.~S6-S8 in SI.
It appears that the migration of the protein to the surface is enhanced by the 
oblate-prolate breathing motion of the droplet. However, these large fluctuations 
may occur because the protein is already charged. The motion of the ions
becomes correlated because there is not enough water for screening the
ions in this high charge density system.
It is observed that the relatively hydrophobic part of the protein extruded first, while the 
hydrophilic and charged part of the protein stayed in the droplet interior and dried-out
as shown in Fig.~\ref{fig:Xvstime}~(c). 
A close-up of the extruded hydrophobic portion is shown in Fig.~\ref{fig:extrusion}. 
At the onset of the extrusion, the droplet stretches into an oval shape. 
As the protein is partially released from the droplet, the droplet relaxes back 
into a spherical shape (Fig.S6 in SI). 
The protein did not completely eject but it dried out (Fig.~S7-S8 in SI).

The same simulation outcome was observed with a different initial condition
where 
a coiled 1WLA$^{22+}$ conformation was placed on the droplet surface, within the 
electric double layer of the droplet.
The possible outcomes of this initial condition are (a) 
protein diffusion towards the interior and (b) 
protein extrusion from the droplet. Since the protein starts from a conformation that 
has been relaxed for the
gaseous state it is bound to extrude. 
From both initial conditions, only a portion of the protein was extruded, while
the rest of the protein dried-out of solvent. 

For a highly charged protein such as the ones used here, a partial extrusion
from small nanodroplets may be viewed as the formation of a \textit{Taylor cone}. In previous research
we have demonstrated the significance of the conical fluctuations 
on the droplet surface for the release of ions\cite{consta2022atomistic, kwan2022conical, kwan2021relation}. 
We also note that in all the simulations of the peptides and myoglobin
the charge is already present in their backbone.
If initially the protein had a low charge then it would have a more organized structure. 
Since the outer surface layers of a droplet holds higher simple ion (e.g. \ce{(H3O+}) concentration than
the bulk interior\cite{consta2013acid}, the protein is subject to a lower pH (see Fig.~2(b)). 
The charging on the surface, may lead to its conformational changes. 
These conformational changes may require much longer time than the simulation time.
Performing the simulations using gas phase equilibrated protein structures
\textit{biases the simulation outcome} since the natural states are delivered from the
bulk solution.

It is emphasized here that these simulations are under non-equilibrium conditions. 
Consequently, the collection of sufficient statistics for such large systems is a formidable tasks. 
The simulations presented here along with those reported by Konermann et al. on myoglobin 
complete ejection\cite{metwally2018chain}  
demonstrate the 
variability of the results depending on the initially selected protein conformation
and its location in the droplet.
A highly charged chain equilibrated for the
gaseous state, is bound to extrude or possibly to be completely ejected because the conformation is
prepared to provide this outcome. Moreover, every highly charged protein placed near the surface will
extrude because of Coulomb repulsion within the chain.

\section{Conclusion}

We used molecular dynamics to address two questions: 
(a) The likelihood of extrusion of poly(ethylene glycol) (PEG), 
linear protonated peptides and protonated  
myoglobin from an aqueous charged droplet. (b) Obtain insight on how  macroion charging  
takes place in droplets larger than those that can be modeled at the atomistic
level. PEG (and like macromolecules) is the only
molecule for which we can directly observe its charging by metal ions and its
release from a charged aqueous droplet. 

We have found that PEG molecules with length between 
54 to $\approx 100$ monomers in droplets composed
up to several thousands of \ce{H2O} molecules, become sodiated and extrude from the droplet.  
The longer chains extrude partially, while the shortest ones (with 54 and 67 monomers), depending
on the temperature, 
may be completely released\cite{constapeg2015, constapeg1, oh2017charging}.
However, the atomistic modeling is limited to a narrow window of droplet sizes in the latest
stages of a droplet's lifetime. In order to examine whether PEG can be charged in
droplets larger than those that can be modeled atomistically we scaled down 
the problem to a large droplet size relative to the length of PEG. 
The charge state of the droplet was near the Rayleigh limit.
We found that PEG54 in a droplet composed of $10^4$\ce{H2O} molecules (equimolar radius
of the droplet 4.2~nm) and
\ce{Na+} ions is frequently sodiated in a transient manner in contrast to 
smaller droplets where long-living (permanent) sodiation takes place. The  
permanent sodiation may lead to the detachment of sodiated PEG from the body of the 
droplet. Finding the existence of  a critical droplet size that affects the
PEG conformation and its charging and by extension that of other macorion, is significant,
as it contradicts the intuitive belief that the more extended the macroion is on a droplet
surface, the higher its charge state would be.

To summarize, there are several factors that affect the charge state of PEG, and by extension
that of other macromolecules such as proteins.
Firstly, the droplet's shape fluctuations.
The initiation of the charging of PEG and the extrusion of the first segment 
takes place 10\%-15\% below the
droplet's Rayleigh limit. At this degree of deviation from 
the Rayleigh limit the droplet's shape fluctuations are
the largest due to charge and the release of the ions is facilitated. 
Secondly, the
temperature. Finally, a new factor was identified in this
study, the droplet's curvature. 
The example of PEG54 in a droplet of $10^4$\ce{H2O} molecules shows that
even though \ce{Na+} ions are available to be captured by PEG, the stretching of the chain
on the surface due to the droplet's curvature prevents the folding of PEG around the ions.
The relative size of length of PEG and droplet curvature is a factor that determines when
long-living chain charging is feasible. Longer chains will be charged and possibly extrude in
larger droplets that those that can be atomistically modeled.

Fenn's ion desorption model (IDM) for linear macroion charging assumes that the free
charges are distributed on the droplet surface at equal distance in order to
minimize the electrostatic repulsion. Fenn suggested that the spacing of charges
will decrease as a droplet shrinks, thus the analyte will be able to capture more
charge.
The availability of free ions in relation to the coverage of the droplet 
surface by a macroion
is definitely a factor that determines the charge state of a macroion. 
However, one has to consider a dynamic location of ions in a thick outer
layer\cite{kwan2020bridging} of a droplet (that includes the electric double layer) 
than a static one.
The present study complements Fenn's intuitive mechanism
by showing that it is not only the free ion availability
but also the droplet's curvature that affects the
conformation of the macroion, and consequently its charging.
Thus, a more stretched macroion on a droplet's surface (that maximizes
the coverage of the surface and thus, is more accessible to
the ions) does not necessarily mean a higher charge
state of the macroion or more facile charging. 

Since protonation reactions cannot be explicitly modeled for proteins in
these systems, the charging of PEG may offer possible scenarios that protons may
be transferred to proteins. The CEM maps the mechanism of charging of PEG 
to unstructured proteins by replacing \ce{Na+} with a hypothetical proton.
However, for proteins it is still unknown at what droplet size the charging may
take place and whether transient protonation plays a role. 
We suggest that it is possible charging to
take place in the low pH region near the
surface at larger droplets than those that we can
model atomistically\cite{consta2013acid}. 

The peptides and myoglobin studied here were already carrying charge when simulated. 
This study revealed that:
(a) Conformations of proteins taken from the gaseous state as it is often done in simulations
of droplets, bias the
simulation outcomes. (b) In the very small nanodroplets, possible ejection of proteins
appears with low probability. (c) Partial extrusion of a protein, when it occurs, may be
viewed as a Taylor cone. 

In previous work we demonstrated the significance of the conical fluctuations on
the droplet surface in the IEM mechanism\cite{sharawy2014effect, kwan2022conical}. 
Similarly, simulations indicate that a rare combination of events 
involving a conical shape deformation may play a role in
a protonated peptide's extrusion. 
It is emphasized that at the Rayleigh limit the
conical deformation appears first and then charged species may enter the cone.
The role of a conical deformation may be that (a) it captures a protein's segment and
assist its extrusion from the droplet surface; 
(b) it allows for simple co-ions (e.g. \ce{H3O+} or \ce{Na+}) in the vicinity of a 
protonated peptide to enter the cone and repel portion of the chain. This step  may assist the
chain to detach from a droplet. Possibly, the conical deformations may
also assist in the charging of macroions. Understanding the chemistry in conical
shapes is still an open question. The role of the conical fluctuations
is robust because we have consistently 
observed them to play a critical role in IEM and here, in different systems containing
the protonated peptide.

Atomistic modeling of droplets with macroions has taken place in minute nanodroplets, in the
regime where a train of single ions cannot be released via Rayleigh jets. 
By simulating this narrow size regime, all the possible pathways of macroion
release such as capturing of macroions in jets cannot be revealed.
Thus, a question arises whether a possible
protein extrusion mechanism found  
at the end of a droplet's lifetime 
is part of the natural path of a droplet's evolution.
It is also noted that detection by simulations of a possible protein extrusion or ejection
will be very sensitive to force field parameters\cite{zapletal2020choice}. 

\section*{Supplementary Material}
(S1) Relaxation of $\rm 1MBN^{17+}$ in an aqueous droplet; 
(S2) Additional data for the charging and conformations of PEG;
(S3) Typical snapshots of the rare event of 
valine-lysine protonated peptides extrusion from charged aqueous droplets. 
(S4) Typical snapshots of $\rm 1WLA^{22+}$[A] partial extrusion and
drying-out in an aqueous droplet. In addition a movie is provided that
shows the transient charging and extrusion of PEG54 in a droplet of $10^4$
and $5\times 10^3$ \ce{H2O} molecules, respectively.

\section*{Acknowledgments}
S.C. is grateful to Prof. D. Frenkel, Yusuf Hamied Department of Chemistry, University of Cambridge, UK 
and Dr. Anatoly
Malevanets for insightful discussions on the stability of charged systems.
S.C thanks Prof. Dr. T. Leisner and Dr. T. Dresch, Karlsruhe Institute of Technology,
Germany, for discussions on the jets and droplet acidity. 
VK and SC acknowledge discussions with Dr. Myong In Oh (SC group) and Dr. Mahmoud Sharawy (SC group), 
who were former graduate students
in the Department of Chemistry, The University of Western Ontario. 
PB and Titiksha acknowledge a MITACS-Globalink funded internship in the Consta lab.
S.C.\ acknowledges an NSERC-Discovery grant (Canada) for funding this
research. V.K. acknowledges the Province of Ontario and The
University of Western Ontario for the Queen Elizabeth II Graduate Scholarship
in Science and Technology. Digital Research Alliance of Canada is acknowledged for
providing the computing facilities.

\providecommand{\latin}[1]{#1}
\makeatletter
\providecommand{\doi}
  {\begingroup\let\do\@makeother\dospecials
  \catcode`\{=1 \catcode`\}=2 \doi@aux}
\providecommand{\doi@aux}[1]{\endgroup\texttt{#1}}
\makeatother
\providecommand*\mcitethebibliography{\thebibliography}
\csname @ifundefined\endcsname{endmcitethebibliography}
  {\let\endmcitethebibliography\endthebibliography}{}

\end{document}